\documentclass[aps,showpacs,preprint,superscriptaddress]{revtex4}
\usepackage{graphicx}
\usepackage{subfigure}
\usepackage{float}
\usepackage{amsmath}
\usepackage{amsfonts}
\usepackage{color}
\usepackage{bm}
\usepackage{bbm}
\usepackage{txfonts}
\usepackage{array}
\usepackage{amssymb}
\begin{document}
\title{Enhanced dynamically assisted pair production in spatial inhomogeneous electric fields with the frequency chirping}

\author{Lie-Juan Li}
\affiliation{Key Laboratory of Beam Technology of the Ministry of Education, and College of Nuclear Science and Technology, Beijing Normal University, Beijing 100875, China}
\author{Melike Mohamedsedik}
\affiliation{Key Laboratory of Beam Technology of the Ministry of Education, and College of Nuclear Science and Technology, Beijing Normal University, Beijing 100875, China}
\author{Bai-Song Xie \footnote{bsxie@bnu.edu.cn}}
\affiliation{Key Laboratory of Beam Technology of the Ministry of Education, and College of Nuclear Science and Technology, Beijing Normal University, Beijing 100875, China}
\affiliation{Beijing Radiation Center, Beijing 100875, China}
\date{\today}

\begin{abstract}
Enhanced dynamically assisted pair production in frequency chirped spatially inhomogeneous electric fields is studied using the Dirac-Heisenberg-Wigner formalism.
The effects of the chirp parameter on the reduced momentum spectrum and the total particle yield are investigated in detail for the single field or the dynamically assisted combined fields at various spatial scales.
For the single high frequency field and the combined fields, the interference effect of momentum spectrum becomes more and more remarkable with the increasing chirp, meanwhile, in the chirped dynamically assisted case, the reduced total yield is enhanced significantly when chirping of two fields are present.
Specifically, in the assisted case, the reduced particle number is increased by more than one order of magnitude at the small spatial scale, while it is enhanced about two times at the large spatial scale.
We also obtain some optimal chirp and spatial scale parameters for the total yield and the enhancement factor for various scenarios.

\end{abstract}
\maketitle

\section{Introduction}

Dirac's equation predicted the existence of positrons \cite {Dirac:1928ej, Anderson:1933mb} and
implied the virtual particles fluctuation inherent in the quantum electrodynamics (QED) vacuum which
further unveiled rich and novel physical phenomena \cite {Dunne:2012vv}.
Sauter noticed that the QED vacuum could produce electron-positron ($e^{+}e^{-}$) pairs
through tunneling at certain field strength scales \cite {Sauter:1931zz} and
Schwinger calculated the pair production rate $\Gamma \sim \exp (-\pi E_{cr}/E)$ for a
constant electric field with proper-time treatment \cite{Schwinger:1951nm}, where
$E_{cr}=m^{2} c^{3} / e \hbar \approx 1.3 \times 10 ^{16} \rm {V/cm}$
is the Schwinger critical field strength corresponding to the laser intensity of
$ I_{cr} \approx 4.3 \times 10^{29} \rm W / \rm cm^{2}$.
Since then, pair production under intense external fields, also known as the Sauter-Schwinger effect,
has become a challenge for both theorists and experimentalists because of its nonperturbative tunneling nature
and the extremely large field strength required for the observation \cite{Gelis:2015kya, DiPiazza:2011tq}.

Fortunately, the development of ultraintense laser technology \cite{Heinzl:2008an, Marklund:2008gj, Pike:2014wha} might make the
experimental observation of pair production possible in the near future.
Especially the application of chirped pulse amplification (CPA) technique has greatly improved the
laser intensity \cite{Strickland:1985gxr},
for example, the current laser intensity has reached about $10 ^{22} \rm W /\rm cm^{2}$ \cite{2008Ultra},
and the expected intensity to be $10 ^{25} \sim 10 ^ {26}\rm W /\rm cm^{2}$ for many
planned facilities \cite{Ringwald:2001ib, Dunne:2008kc}.
On the other hand, the X-ray free electron laser (XFEL) system may achieve subcritical field strength as large as
$E\approx 0.01-0.1 E_{cr}$ \cite{Ringwald:2001ib} so that the vacuum $e^{+}e^{-}$
pair production has become a hot research topic in past decade.
So far, several methods have been developed to investigate pair production in external fields,
such as the worldline instanton technique \cite {Gies:2005bz, Dunne:2006st, Schneider:2014mla},
quantum Vlasov equation solution method \cite {Kluger:1991ib, Alkofer:2001ik, Abdukerim:2017hkh},
computational quantum field theory \cite{P2005Creation,PhysRevLett.111.183204,Wang:2019oyk},
Dirac-Heisenberg-Wigner (DHW)
formalism \cite{Hebenstreit:2011wk, Kohlfurst:2015zxi, Xie:re2017, Kohlfurst:2017hbd, Ababekri:2019qiw,Kohlf2020Effect},
and so on.

Theoretically, many different field configuration, {e.g.}, the temporal and spatial pulse shaping, have been investigated to reduce the pair creation threshold.
The research on pair production in short laser pulses with subcycle structure discovered that the
momentum spectrum is extremely sensitive to the corresponding external field parameters \cite{Hebenstreit:2009km}.
Particle self-bunching effect is identified in pair creation under the spatially inhomogeneous pulse field \cite{Hebenstreit:2011wk}.
The ponderomotive effect is found important in the multiphoton pair production process for temporal oscillating and spatially
inhomogeneous electric field \cite {Kohlfurst:2017hbd}.
As an important step to increase the pair production, Sch\"{u}tzhold  \textsl{et al}. \cite{Schutzhold:2008pz} proposed dynamically assisted Schwinger mechanism, which is the combination of a low frequency strong field with a high frequency weak field. They
pointed out that the pair creation rate is enhanced extremely by the combined field effect.
Also, frequency chirp effects are widely studied because they affect the momentum spectra and
the total particle number under the time dependent field \cite{Dumlu:2010vv,Olugh:2018seh}
and spatial inhomogeneous oscillating electric field \cite {Ababekri:2019qiw}.
Furthermore, many different combinational field have been studied to enhance the $e^{+}e^{-}$ pair
production \cite{Orthaber:2011cm,Ababekri:2019dkl, Olugh:2019nej}.
Especially, the enhanced pair production in time dependent one- and two-color laser fields
by chirping have also been explored \cite{Abdukerim:2017hkh, Gong:2019sbw}.

In this paper, we use the DHW formalism to investigate pair production in frequency chirped dynamically assisted electric field with spatial inhomogeneity. We research the reduced momentum spectrum, the reduced total yield of the created pairs for either low or
high frequency one-color field with different chirp values. Meanwhile, we study chirped dynamically assisted pair production for the two-color combined fields. Many different momentum spectrum and pair production number as well as the enhanced factor are presented for various chirping cases.
Throughout this paper natural units $\hbar=c=1$ are used and all quantities are expressed in terms of
the electron mass $m$.

The paper is organized as follows. In Sec. \ref{method},
we introduce the background field that is considered in this work, and
review the DHW formalism briefly.
In Sec. \ref{single external field}, we show the numerical results for the single
electric field with different chirp values. In Sec. \ref{DSM}, we present
the numerical results for chirped dynamically assisted two-color combined fields and semiclassical discussion
about the results.
In Sec. \ref{conclusion}, we give a brief conclusion and outlook.

\section{Background field and theoretical formalism}\label{method}

\subsection{Model of the background electric field}\label{field}

We investigate $e^{+}e^{-}$ pair production
for either low or high frequency one-color field and two-color dynamically assisted combinational fields with various chirp parameters.
Our idealized model is considered as the combination of a low frequency and strong field $E_{1s}\left(x,t\right)$
with a high frequency and weak field $E_{2w}\left(x,t\right)$. The form is given as
\begin{equation}\label{FieldMode}
\begin{aligned}
E\left(x,t\right)
&= E_{1s}\left(x,t\right)+E_{2w}\left(x,t\right)\\
&=E_{1s0}\exp \left(-\frac{x^{2}}{2 \lambda^{2}} \right )\cos^{4}\left(\frac{t}{\tau}\right)\cos(b_{1} t^{2}+\omega_{1} t) \\
&+E_{2w0}\exp \left(-\frac{x^{2}}{2 \lambda^{2}} \right )\cos^{4}\left(\frac{t}{\tau}\right) \cos(b_{2} t^{2}+\omega_{2} t),
\end{aligned}
\end{equation}
where $E_{1s0,2w0}$ denote the strength of the electric field, $b_{1,2}$ are the corresponding frequency chirp,
$\omega_{1}$ and $\omega_{2}$ are the oscillating frequency of the strong and weak fields,
$\tau$ and $\lambda$ are the characteristic time and spatial length scale, respectively.
Note that the temporal pulse shape function $\cos^{4}(t/\tau)$ is turned on at $t=-\pi \tau/2$
and turned off at $t=\pi \tau/2$. This pulse duration condition is applied to all the cases of studied fields in this paper.

In our numerical calculation, a set of parameters of the electric field Eq. (\ref{FieldMode}) are fixed as
\begin{equation}\label{Fixed parameters}
\begin{aligned}
E_{1s0} = 0.3E_{cr}, E_{2w0} = 0.075 E_{cr}=0.25E_{1s0},\\
\omega_{1} = 0.1m, \omega_{2} = 0.7m=7\omega_{1}, \tau = 50m^{-1},
\end{aligned}
\end{equation}
where $E_{cr}$ is the critical field strength and $m$ the electron mass. Note that the Keldysh parameter is
important for studying the pair creation from vacuum and defined as $\gamma = m\omega/eE$ \cite{Keldysh:1965ojf,Brezin:1970xf},
where $E$ and $\omega$ represent the strength and frequency of the background field, and the Schwinger (tunneling)
effect and multiphoton absorption can be considered by $\gamma \ll 1$ and $\gamma\gg 1$.
For the chosen parameters in this paper, the Keldysh parameter $\gamma_{1s}=m\omega_{1}/eE_{1s0}= 1/3$
of the single field $E_{1s}\left(x,t\right)$ means that the pair production process is dominated by tunneling effect whereas
$\gamma_{2w}=m\omega_{2}/eE_{2w0}=28/3$ of the field $E_{2w}\left(x,t\right)$ means that the process is dominated by multiphoton absorption.
It must be noted that the combined Keldysh parameter $\gamma_{C} = m\omega_{2}/eE_{1s0}$=7/3 in this study
is the combination of the strong field strength $E_{1s0}$ and the fast
pulse frequency $\omega_{2}$ \cite{Schutzhold:2008pz}, i.e., the process of pair creation involve both
multiphoton absorption and tunneling effect.
We also use the definition of the time-dependent effective frequency \cite{Ababekri:2019qiw} $\omega_{\text{eff}}(t)=\omega+bt$, with $b$ denoting the frequency chirp parameter.
Since that the pair creation process occurs mostly at the dominate time interval $-\tau \le t \le \tau$, so
the chirp parameter in the "normal regime", i.e., $\max{b|t|} \sim\omega$ in this interval must be met naturally. Therefore,
in this paper, the values of linear chirp $b$ can be expressed as $b=\alpha \omega/\tau$ with $0< \alpha < 1$.
Without losing the generality, we choose the regime of $0\leq \alpha \leq 0.9$ so that the studied maximum chirp value is chosen as $b_{1}=0.9 \omega_{1}/\tau = 0.0018 m^2$ for $E_{1s}\left(x,t\right)$. Similarly for $E_{2w}\left(x,t\right)$, the maximum chirp value is chosen as $b_{2}=0.9\omega_{2}/\tau = 0.0126 m^2$.
Furthermore, for the simplicity of study and convenience of results comparison, we limit the selection rule for chirping as that the ratio of frequency chirp value of two fields keeps the same ratio of the original center frequency of two fields, i.e., $b_1/b_2=\omega_1/\omega_2$.

Note that the electric field pulse shape and also the chosen field parameters are different from that of our previous research \cite{Ababekri:2019qiw}.
Beside obtaining the rich momentum spectrum, total yield of pair production and enhancement factor in different fields and chirping cases, moreover, we have
explored some optimal chirp parameters and spatial scales for $e^{+}e^{-}$ pair production via extensive numerical simulation.

\subsection{Theoretical formalism: DHW formalism}\label{DHWformalism}

DHW formalism, a relativistic phase-space quantum kinetic approach \cite{Vasak:1987um,Phase-space str}, which has been widely used to explore pair production within arbitrary electromagnetic \cite{Hebenstreit:2011wk,Olugh:2018seh,Ababekri:2019dkl,Kohlf2020Effect}. As the specific derivation of DHW has been given in  previous works \cite{Phase-space str,Kohlfurst:2015zxi} and the detailed introduction is not the purpose of this work, we only present the basic ideas and the essential points of this method.

We begin with the gauge-covariant density operator of the system based on \cite{Phase-space str}
\begin{equation}\label{DensityOperator}
 \hat {\mathcal C}_{\alpha \beta} \left(r , s \right) = \mathcal U \left(A,r,s
\right) \ \left[ \bar \psi_\beta \left( r - s/2 \right), \psi_\alpha \left( r +
s/2 \right) \right],
\end{equation}
where $r$ represents the center-of-mass coordinate and $s$ the relative coordinate. The Wilson line factor $\mathcal U \left(A,r,s \right)$ before the commutator
\begin{equation}\label{Wilson line factor}
 \mathcal U \left(A,r,s \right) = \exp \left( \mathrm{i} \ e \ s \int_{-1/2}^{1/2} d
\xi \ A \left(r+ \xi s \right)  \right),
\end{equation}
is a factor which can guarantee the gauge invariance.
Obviously, it is related to the elementary charge $e$ and the background gauge field $A$.
The covariant Wigner operator is obtained via the Fourier transform of the covariant density operator from the $s$-space to the $p$-space
\begin{equation}\label{WignerOperator}
 \hat{\mathcal W}_{\alpha \beta} \left( r , p \right) = \frac{1}{2} \int d^4 s \
\mathrm{e}^{\mathrm{i} ps} \  \hat{\mathcal C}_{\alpha \beta} \left( r , s
\right).
\end{equation}

In order to make the  equation of motion, expressed in terms of Wigner operator, reasonable for calculation, the background field is treated
as that in mean-field (Hartree) approximation \cite{Kohlf2020Effect}, {i.e.},
\begin{equation}\label{Electromagnetic field tensor}
 F^{\mu \nu} \left( {r} \right) \approx \langle \hat F^{\mu \nu} \left(
{r} \right) \rangle.
\end{equation}
The Hartree approximation becomes apparent when we consider the vacuum expectation value of the covariant Wigner operator to get the Wigner function
\begin{equation}\label{Wigner function}
 \mathbbm{W} \left( r,p \right) = \langle \Phi \vert \hat{\mathcal W} \left( r,p
\right) \vert \Phi \rangle.
\end{equation}
As a result, the electromagnetic field factors out
\begin{equation}\label{electromagnetic field}
 \langle \Phi \vert F_{\mu \nu} \ \hat{\mathcal{C}} \vert \Phi \rangle =
F_{\mu \nu} \langle \Phi \vert \hat{\mathcal{C}}
\vert \Phi \rangle \,.
\end{equation}
Wigner function is in the Dirac algebra, it can be expanded in terms of 16 covariant Wigner coefficients
\begin{equation}\label{decomposed}
\mathbbm{W} = \frac{1}{4} \left( \mathbbm{1} \mathbbm{S} + \textrm{i} \gamma_5
\mathbbm{P} + \gamma^{\mu} \mathbbm{V}_{\mu} + \gamma^{\mu} \gamma_5
\mathbbm{A}_{\mu} + \sigma^{\mu \nu} \mathbbm{T}_{\mu \nu} \right) \,.
\end{equation}

Combined with the above introduction, the equation of motion in terms of covariant Wigner function can be obtained.
However, the numerical solving of the equation of motion from covariant Wigner function at all spacetime points is hardly task to perform, therefore, a convenient and powerful treatment is to obtain the equal-time Wigner function
\begin{align}
 \mathbbm{w} \left( \mathbf{x}, \mathbf{p}, t \right) = \int \frac{d p_0}{2 \pi}
\ \mathbbm{W} \left( r,p \right).
\end{align}

When investigating the $1+1$ dimensional spatial inhomogeneous electric field given in Eq.(\ref{FieldMode}) one finds only
$\mathbbm{S}$, $\mathbbm{P}$, and $\mathbbm{V}_{\mu} (\mu=0,1)$ are nonvanishing, and the corresponding physical quantities are written in lowercase in the
equal-time case. Hence, the equations of motion can be reduced to $4$ equations \cite{Hebenstreit:2011wk,Kohlfurst:2015zxi}
\begin{align}
 &D_t \mathbbm{s} - 2 p_x \mathbbm{p} = 0 , \label{pde:1}\\
 &D_t \mathbbm{v}_{0} + \partial _{x} \mathbbm{v}_{1} = 0 , \label{pde:2}\\
 &D_t \mathbbm{v}_{1} + \partial _{x} \mathbbm{v}_{0} = -2 m \mathbbm{p} , \label{pde:3}\\
 &D_t \mathbbm{p} + 2 p_x \mathbbm{s} = 2 m \mathbbm{v}_{1} , \label{pde:4}
\end{align}
with the differential operator
\begin{equation}\label{pseudoDiff}
 D_t = \partial_{t} + e \int_{-1/2}^{1/2} d \xi \,\,\, E_{x} \left( x + i \xi \partial_{p_{x}} \, , t \right) \partial_{p_{x}} .
\end{equation}
Corresponding vacuum initial conditions are given \cite{Kohlfurst:2015zxi} by
\begin{equation}\label{vacuum-initial}
{\mathbbm s}_{vac} = - \frac{2m}{\omega} \, ,
\quad {\mathbbm v}_{vac} = - \frac{2{ p_x} }{\omega} \,  ,
\end{equation}
where $\omega=\sqrt{p_{x}^{2}+m^2}$ is the energy of a particle.
The particle number density is given \cite{Hebenstreit:2011wk,Kohlfurst:2017hbd}
\begin{equation}\label{PS}
n \left( x , p_{x} , t \right) = \frac{m \mathbbm{s}^{v} \left( x , p_{x} , t \right) + p_{x}  \mathbbm{v}^{v} \left( x , p_{x} , t \right)}{\omega \left( p_{x} \right)},
\end{equation}
in which $ \mathbbm{s}^{v}\left( x , p_{x} , t \right)=\mathbbm {s} - {\mathbbm s}_{vac}$ and
$\mathbbm{v}^{v} \left( x , p_{x} , t \right)=\mathbbm {v} - {\mathbbm v}_{vac}$ are the modified Wigner components.
The momentum space particles number density could be obtained  from $n \left( x , p_{x} , t \right)$  by integrating out $x$
\begin{equation}\label{MS}
n \left( p_{x} , t \right) = \int d x \, \frac{\mathbbm{s}^{v} \left( x , p_{x} , t \right) + p_{x} \, \mathbbm{v}_{1}^{v} \left( x , p_{x} , t \right)}{\omega \left( p_{x} \right)}.
\end{equation}
Accordingly, the total yield of created particles can be written
\begin{equation}\label{Num}
N\left(t \right) = \int dx dp_x n \left( x , p_{x} , t \right).
\end{equation}
We will investigate the reduced quantities including the reduced particle number density $\bar n \left( p_{x} , t \right) \equiv \frac{n \left( p_{x} , t \right) }{\lambda}$ and reduced total yield of the created particles $\bar N\left(t\rightarrow \infty \right) \equiv \frac{N\left(t \rightarrow \infty \right)}{\lambda}$ to obtain the effect of nontrivial spatial scale $\lambda$.

\section{Numerical results for the single field}\label{single external field}

In this section, we consider the influence of chirp values on the pair production for the single field
with respect to either $E_{1s}\left(x,t\right)$ or $E_{2w}\left(x,t\right)$.

\subsection{Single field $E_{1s}\left(x,t\right)$ } \label{result E1s}

\begin{figure}[htbp]\suppressfloats
\includegraphics[scale=0.5]{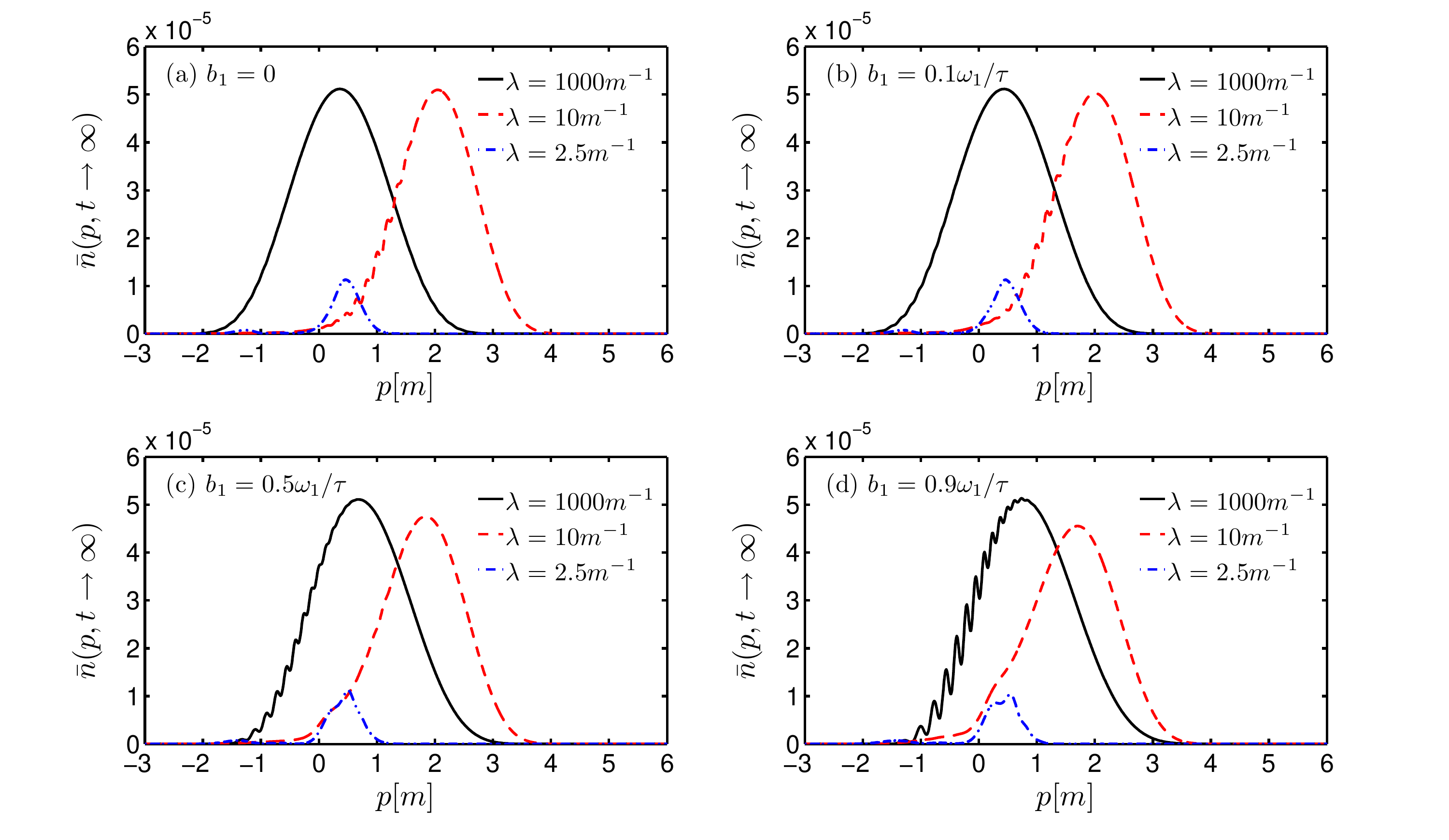}
\caption{(color online). Reduced momentum spectrum for different
spatial scales ($\lambda=1000m^{-1}, 10m^{-1}, 2.5m^{-1}$) for one-color strong laser field
$E_{1s}(x,t)$ with various chirp values of
$b_{1}[\omega_{1}/\tau]$ as $0$, $0.1$, $0.5$ and $0.9$, i.e., $b_{1}[m^2]$
as $0$, $0.0002$, $0.001$ and $0.0018$, from (a) to (d) respectively.
The field parameters are $E_{1s0}=0.3E_{cr}$, $\omega_{1}=0.1m$ and $\tau=50m^{-1}$.
To eyes guidance, we amplify
the value by $10$ times when $\lambda=2.5m^{-1}$.}
\label{fig:1}
\end{figure}

Firstly, we investigate the pair production in the one-color strong field $E_{1s}\left(x,t\right)$,
i.e., the Sauter-Schwinger effect is dominant. In this case, the individual field reads
\begin{equation}\label{E1sMode}
\begin{aligned}
E_{1s}\left(x,t\right)
&=E_{1s0}\exp \left(-\frac{x^{2}}{2 \lambda^{2}} \right )\cos^{4}\left(\frac{t}{\tau}\right)\cos(b_{1} t^{2}+\omega_{1} t).
\end{aligned}
\end{equation}
The meaning of the corresponding parameters has been introduced in detail in section \ref{field}.
The fixed parameters for this field are $E_{1s0} = 0.3E_{cr}$,
$\omega_{1} = 0.1m$, and $\tau = 50m^{-1}$.
The momentum spectrum for different spatial scales with various chirp values of $b_{1}$ is given in Fig. \ref{fig:1}.

As shown in Fig. \ref{fig:1}(a), i.e., for $E_{1s}\left(x,t\right)$ with vanishing chirp $b_{1}=0$,
at the quasihomogeneous limit $\lambda=1000 m^{-1}$,
the center of the reduced momentum spectrum does not corresponding to $p=0$
which is similar to the homogeneous case in Fig. $2$
of the Ref. \cite {Hebenstreit:2009km}. This is because the vector potential
value $A(t)\neq 0$ when the external field is turned off,
so that $p_{x}=q-eA(t)$ has a nonzero value $-A(t\rightarrow \infty)$ corresponding to the momentum peak at $q=0$.
In fact when $b_{1}=0$, and we neglect the spatial variation of field when $\lambda$ is large, then $A(t)=-\int_{0}^{t} E_{1s0}\cos^{4}\left(t/\tau\right)\cos(\omega_{1} t)dt=-(I_{1}+I_{2}+I_{3})$ can be obtained, where
$I_{1}= 3/8 \int_{0}^{t} \cos(5 t/\tau)dt$, $I_{2}=1/2\int_{0}^{t} \cos(2 t/\tau) \cos(5 t/\tau)dt$ and $I_{3}=1/8\int_{0}^{t} \cos(4 t/\tau) \cos(5 t/\tau)dt$. For $E_{1s0}=0.3 E_{cr}$, $\omega_{1}=0.1m$ and $\tau=50m^{-1}$, we have $A(t)\approx-0.381$ when the external field is turned off at $t=\pi \tau/2$, and the kinetic momentum $p_{x}\approx0.381$. This is approximately consistent with the numerical result $p_{x}\approx0.366$ when $\lambda=1000 m^{-1}$ shown in Fig. \ref{fig:1}(a). In the chirping cases with different inhomogenous field scale $\lambda$ in Fig. \ref{fig:1}, the reason of the momentum peak shifting from zero is the same although the momentum values of peak position are different because the $-A(t\rightarrow \infty)$ depend strongly on the $b$ and $\lambda$.
For example, when the spatial scale decreases to $\lambda=10m^{-1}$,  momentum spectrum shifts to the larger momentum values due to the effect
of relatively strong spatial inhomogeneity of the electric field \cite{Ababekri:2019qiw}. This can be understood qualitatively as that the electric field strength decreases rapidly when the spatial scale $\lambda$ is small makes the created particles with high momenta easier to escape the regime of field existence so that they are hardly affected by the field.
By comparing the results shown in Fig. \ref{fig:1}(b) and (c), we can see that momentum spectrum oscillates
significantly with the increase of chirp in the quasihomogeneous case. The phenomenon could be understood as the
interference effect of particles created by opposite signed large peaks of the temporal field \cite{Dumlu:2010vv}.
It can be seen from Fig. \ref{fig:1}(d) that the momentum spectrum oscillation is stronger as the chirp $b_{1}$ increases to a large one because of the enhanced interference effect between created particles. In particular this oscillation behavior is very remarkable for quasihomogenous case when spatial scale $\lambda=1000 m^{-1}$.

On the other hand, by comparing the particle number density of the small spatial scale, e.g. $\lambda=2.5m^{-1}$, with that of the large spatial scale,
it decreases notably when $b$ is fixed. This is because that the work done by the electric field is smaller when spatial scale is small so that the particles created by the electric field are also smaller.
Moreover, one sees that when the chirp is applied, the maximum peak value of the momentum spectra does not increase apparently in all cases. It indicates that the chirp parameter we choose has a little effect on the peak value of momentum spectrum in the case of the strong field $E_{1s}\left(x,t\right)$. There are two reasons for this result. On the one hand, the contribution of strong fields to the particles creation is dominated by the field intensity rather than the frequency, which is usually exponent suppression in subcritical field strength. On the other hand, in the case of low frequency electric field studied here,
even if in the presence of frequency chirping, the frequency spectrum is only extended to two or three times of the original center frequency by the Fourier transform of the electric field Eq. (\ref{E1sMode}). Therefore, the possible dynamically assisted mechanism is suppressed strongly. Because we know that the dynamically assisted mechanism is expected to play a key role only when the chirping leads to about one-order magnitude increasing for frequency component compare to the original frequency.

In order to understand the momentum spectra behavior, we shall make some discussions on our numerical results qualitatively by using Wentzel-Kramers-Brillouin (WKB) method \cite{Dumlu:2010ua} in Sec. \ref{SemiDis} in the following. The turning points are presented in Fig.\ref{fig:15} and Fig.\ref{fig:16} for typical fields we studied in this paper. Note that the plotting of turning points in case of low frequency strong field studied in this subsection is given in Fig.\ref{fig:15}.

\subsection{Single field $E_{2w}\left(x,t\right)$}\label{result E2w}

In this subsection, we investigate the momentum spectrum and total yield
in one-color weak field $E_{2w}\left(x,t\right)$ with various chirp values of $b_{2}$.
In this case, the $n$-photon pair production is dominant and the model of the external electric field is

\begin{equation}\label{E2wMode}
\begin{aligned}
E\left(x,t\right)
&=E_{2w0}\exp \left(-\frac{x^{2}}{2 \lambda^{2}} \right )\cos^{4}\left(\frac{t}{\tau}\right) \cos(b_{2} t^{2}+\omega_{2} t),
\end{aligned}
\end{equation}
where the calculation parameters are chosen as $E_{2w0} = 0.075 E_{cr}$,
$\omega_{2} = 0.7m$ and $\tau = 50m^{-1}$.

\begin{figure}[htbp]\suppressfloats
\includegraphics[scale=0.5]{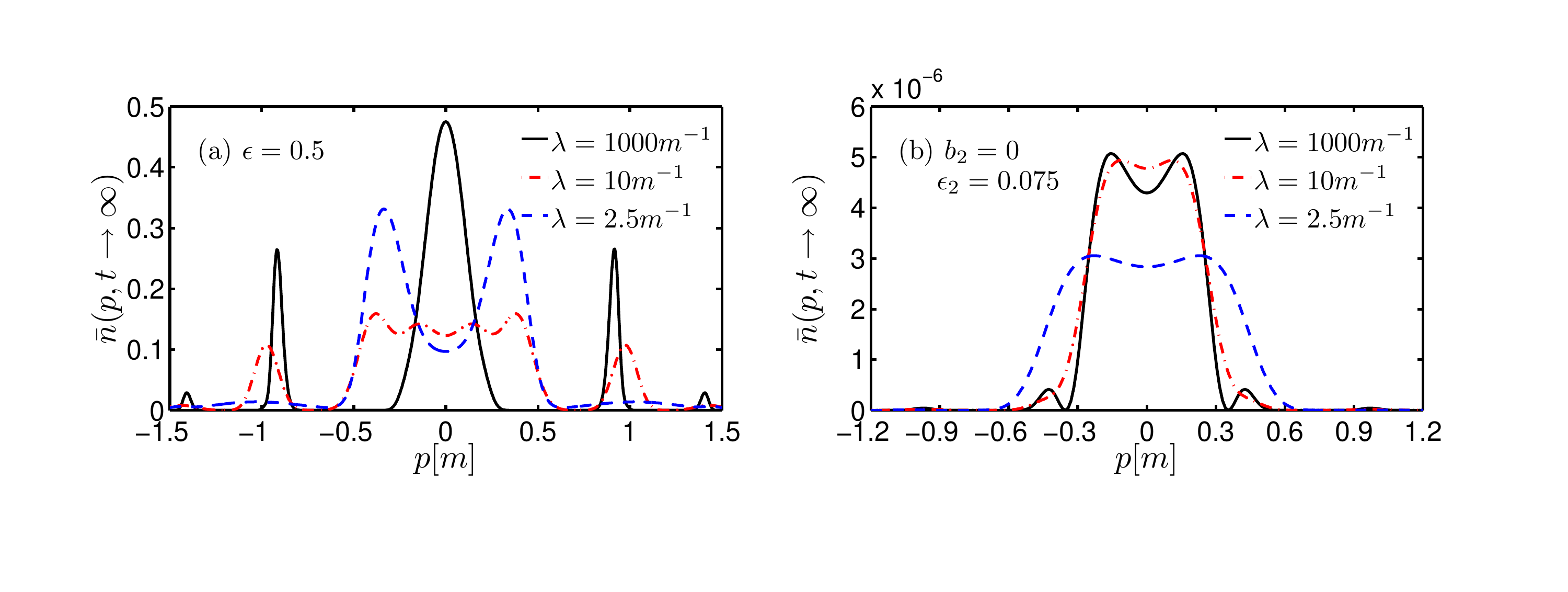}
\caption{(color online). Comparison of reduced momentum spectrum for different spatial scales
between the results in Ref. \cite{Kohlfurst:2017hbd} and our results.
The left one is the Fig. 3 in Ref. \cite{Kohlfurst:2017hbd} with field parameters $\epsilon=0.5$,
$\tau=100m^{-1}$ and $\omega=0.7m$.
The right one is our results for one-color weak laser $E_{2w}\left(x,t\right)$ without chirping
with field parameters $E_{2w0}=0.075E_{cr}$, $\omega_2=0.7m$ and $\tau=50m^{-1}$.}
\label{fig:2}
\end{figure}

In order to guarantee the reasonability of the chosen parameters, we reproduce the results of
the Fig.3 in Ref. \cite{Kohlfurst:2017hbd}
as shown in Fig. \ref{fig:2}(a), where the parameters are chosen as field strength $\epsilon=0.5$,
the envelop pulse length $\tau=100m^{-1}$ and frequency $\omega=0.7m$.
The reduced particle momentum spectrum for one-color weak
laser $E_{2w}\left(x,t\right)$ with vanishing chirp $b_{2}=0$
for the different spatial scales in our studied case is shown in Fig. \ref{fig:2}(b).
Correspondingly, the strength of our field $E_{2w0}$ =$\epsilon_{2}E_{cr}$
with $\epsilon_{2}=0.075$ that is appear in the Fig. \ref{fig:2}(b).

We can see that there is a difference of $5$ orders of magnitude between the peak
value of $\bar n$ in Fig. \ref{fig:2}(a) and that in Fig. \ref{fig:2}(b).
Estimating the value through the relationship between the particle number density
$\bar n \left( p_{x} , t \right)$ of created particles and the strength $\epsilon_{2}$ of $E_{2w}\left(x,t\right)$.
The expression is $\bar n \propto \epsilon_{2}^{2N_{p}}$, where $N_{p}=3$ denotes
the number of photon during the process of multipohton absorption.
Thus, we get $(\epsilon_2/\epsilon)^6 = (0.075/0.5)^6 \approx 1.139 \times 10^{-5}$, which is
consistent with numerical results of our check example in Fig. \ref{fig:2} comparable to Fig.3 in Ref. \cite{Kohlfurst:2017hbd}.
By the way, this estimation can be as a test of the correctness of the numerical results performed in this paper.

From Fig. \ref{fig:2}(b),
one can see that the peak splitting on the momentum spectrum at the large spatial
scale $\lambda=1000 m^{-1}$ because of the interference
effect \cite{Dumlu:2011rr}. The interference effect is caused by the superposition of wavefunctions
when particles experience different turning points due to the applied field, i.e., physically this effect is produced by the waves reflected by the multi-bump structure so that there has an additional phase shift on reflection. To understand this more in detail, we have depicted the turning point structure in Fig. \ref{fig:16}, along with some discussions in Sec. \ref{SemiDis}.

On the other hand, at the small spatial scales $\lambda =10 m^{-1}$ and $\lambda =2.5 m^{-1}$,
the observed peak splitting on momentum spectrum can be understood in terms of ponderomotive forces which is
beneficial for the production of particles by pushing particles out of production region \cite{Kohlfurst:2017hbd}. Note that the strength of ponderomotive forces is
inversely proportional to the size of spatial scale $\lambda$,
that is, the smaller the $\lambda$, the more obvious the pushing of the ponderomotive forces.
Consequently, the momentum peaks at spatial scale $\lambda =2.5 m^{-1}$ are pushed far away from the center
compared with that at $\lambda =10 m^{-1}$.

\begin{figure}[htbp]\suppressfloats
\includegraphics[scale=0.5]{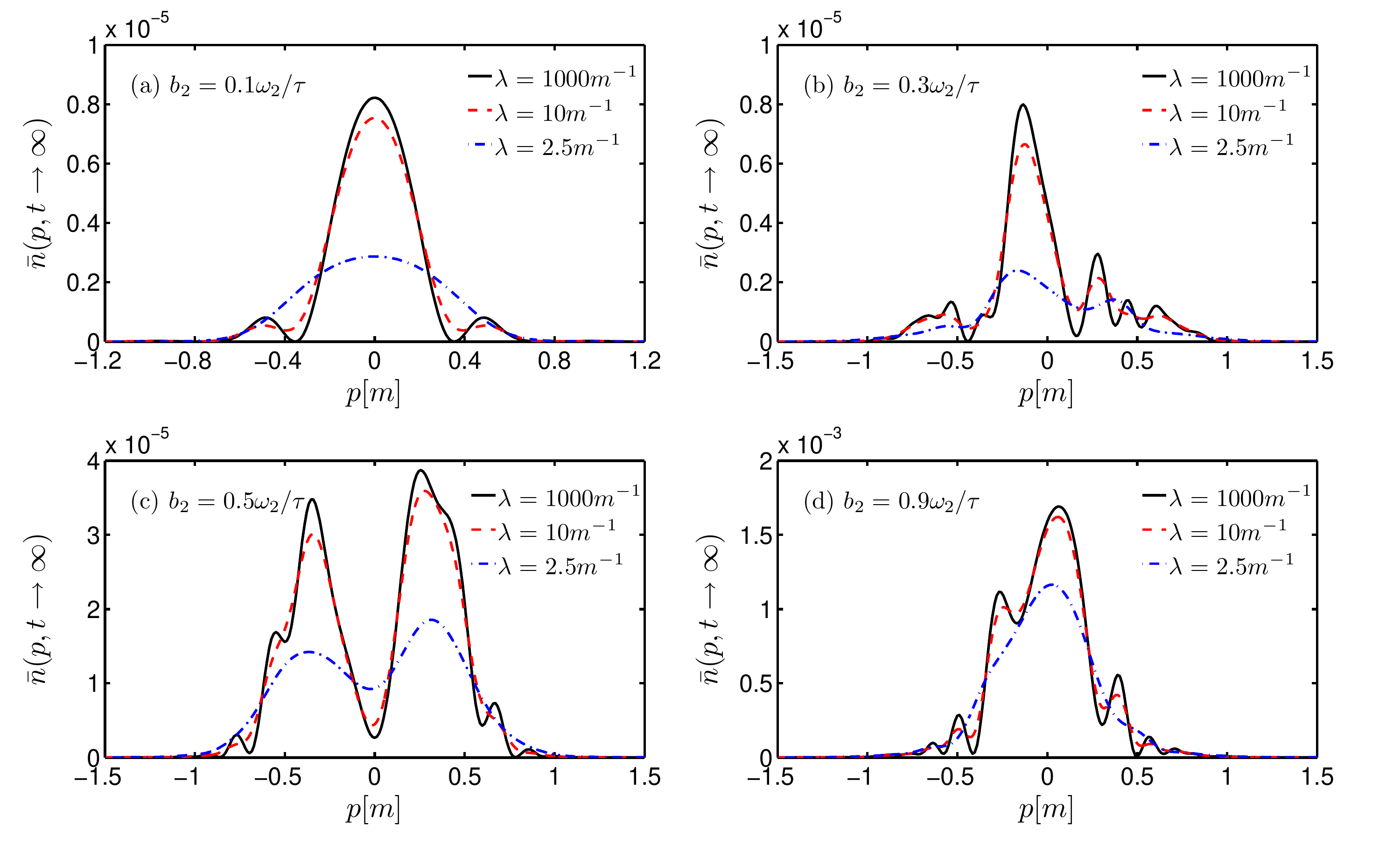}
\caption{(color online). Reduced momentum spectrum of created $e^{+}e^{-}$ pair
for different spatial scales for one-color weak field
$E_{2w}(x,t)$ with different chirp of $b_{2} [\omega_{2}/\tau]$ as $0.1$, $0.3$, $0.5$, $0.9$, i.e.,
$b_{2} [m^{2}]$ as $0.0014$, $0.0042$, $0.007$, $0.0126$, from (a) to (d) respectively.
The other parameters are the same as in Fig. \ref{fig:2}(b).}
\label{fig:3}
\end{figure}

In addition, the momentum spectrum of created $e^{+}e^{-}$ pair
for different spatial scales for one-color weak field
$E_{2w}(x,t)$ with different chirp values of $b_{2}$ is shown in Fig. \ref{fig:3}.
As can be seen in the Fig. \ref{fig:3}(a), for small chirp $b_{2}=0.1 \omega_{2}/\tau$, the main peak located around $p=0$ for different spatial scales. However, there appears also two less pronounced peaks when $\lambda=1000m^{-1}$. In other words, the spectrum is very sensitive to chirps even the value of chirp is relatively small. Meanwhile, the peak is increased compared
with the results without chirping ($b_{2}=0$) shown in
Fig. \ref{fig:2}(b).

For intermediate chirp $ b_{2}=0.3 \omega_{2}/\tau$,
we can see that there is not only one main peak
but also a number of smaller peaks as shown in Fig. \ref{fig:3}(b).
When the chirp increases to $ b_{2}=0.5 \omega_{2}/\tau$,
the peaks become two dominate maxima and also there exist other less pronounced mini-peaks, see Fig. \ref{fig:3}(c).
Also, for the large value of chirp $b_{2}=0.9\omega_{2}/\tau$, the spectrum exhibits incomplete interference,
and the rate becomes higher compared
with the other cases due to the increase in effective frequency of the electric field.

The results show that as the increase of the chirp values, the coherent interference pattern in the momentum spectrum is
changed and it turns into complex oscillatory.
These effects can be qualitatively discussed with the interference between the complex conjugate pairs of turning points.
The number of turning point pairs that are almost equidistant from the real axis increases
with the increase of the chirp value. Therefore, stronger interference effect
can be observed on the momentum spectrum for large chirp. This is confirmed by the typical structure of the turning points
as an illustration in Fig. \ref{fig:16}.

\begin{figure}[htbp]\suppressfloats
\includegraphics[scale=0.42]{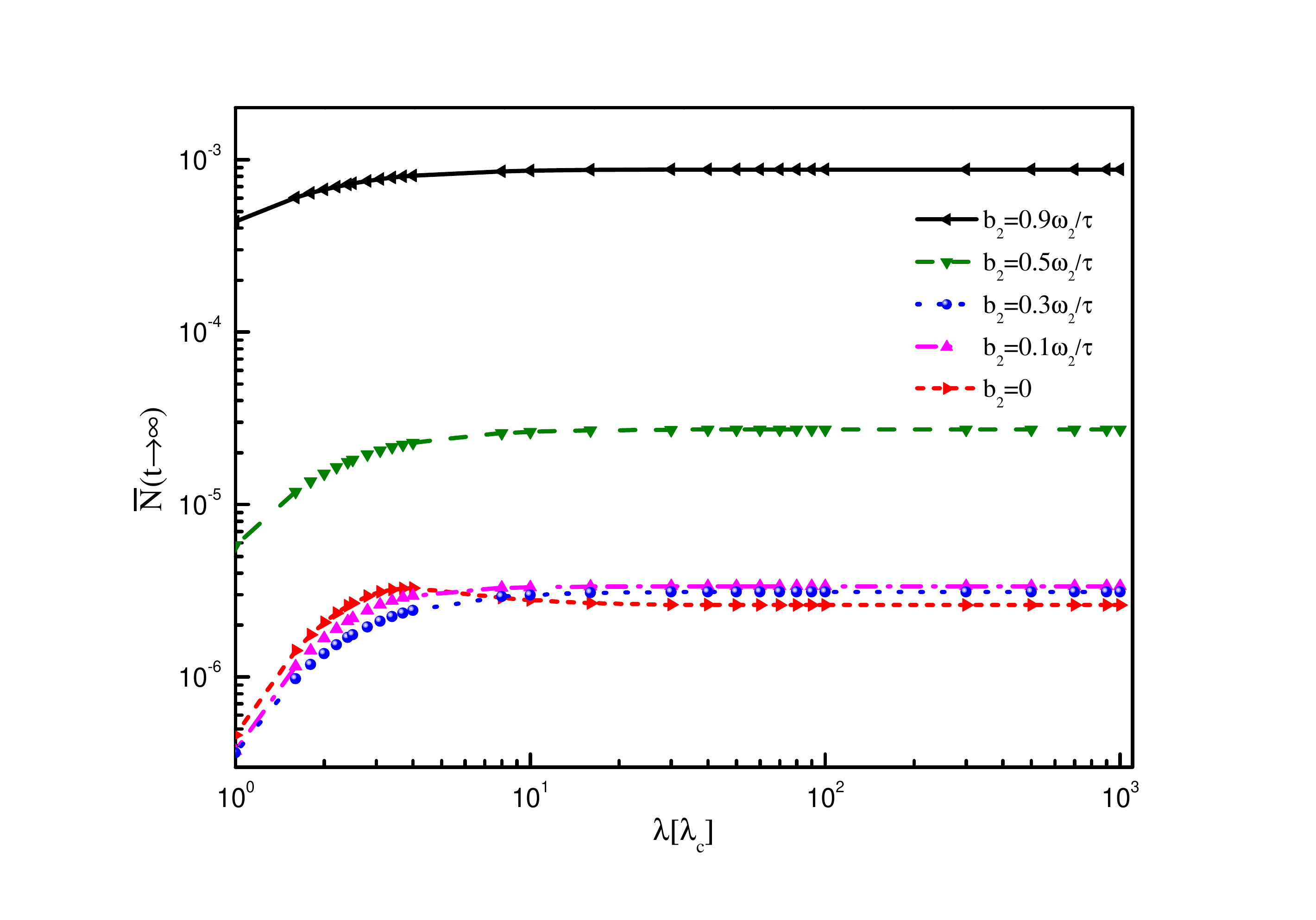}
\caption{(color online). Reduced total yield $\bar N (t \rightarrow \infty )$ for $E_{2w}\left(x,t\right)$ with
different chirp as a function of the spatial scales. The other parameters are the same as in Fig. \ref{fig:3}.}
\label{fig:4}
\end{figure}

In order to investigate the influence of the chirp values more deeply, the reduced total yield $\bar N (t \rightarrow \infty )$ for $E_{2w}\left(x,t\right)$ with different chirp values as a function of the spatial scale is displayed in Fig. \ref{fig:4}.
We can see that the particle number is enhanced with the increasing of chirp value.
For the small chirp values, at least for $b_{2} \le 0.3 \omega_{2}/\tau$, the total yield of created particles quickly decreases
with the shrinking of spatial scale in finite spatial scales.
Because the electric field energy decreases with decreasing of spatial scale, particles
created in the field region also decrease correspondingly.
For the intermediate chirp $b_{2}=0.5\omega_{2}/\tau$, the reduced total yield drops slowly with decreasing spatial scale. For the large chirp $b_{2}=0.9\omega_{2}/\tau$, $\bar N (t\rightarrow \infty )$ tends to an almost flat line, because the particles creation process is mainly high energy photon absorption that is less affected by the spatial scale.

\section{Numerical results for two-color combinational fields }\label{DSM}

In this section, we turn to investigate the numerical results of inhomogeneous dynamically assisted
combinational electric fields with several chirped forms. The model of the field is shown in Eq. (\ref{FieldMode})
and the fixed field parameters are given in Eq. (\ref{Fixed parameters}).
We explore the reduced momentum spectrum or/and the reduced total yield of created particles
as well as enhancement factor for several spatial scales, for the following cases:
(i) without chirp ($b_{1}=b_{2}=0$),
(ii) chirping is only for $E_{1s}\left(x,t\right)$ ($b_{1} \neq 0, b_{2}=0$),
(iii) chirping is only for $E_{2w}\left(x,t\right)$ ($b_{1}=0, b_{2} \neq 0$),
and (iv) chirping is for both $E_{1s}\left(x,t\right)$ and $E_{2w}\left(x,t\right)$ ($b_{1} \neq 0,b_{2} \neq 0$).

\subsection{Two-color field without chirping}\label{result1b}

\begin{figure}[htbp]\suppressfloats
\includegraphics[scale=0.5]{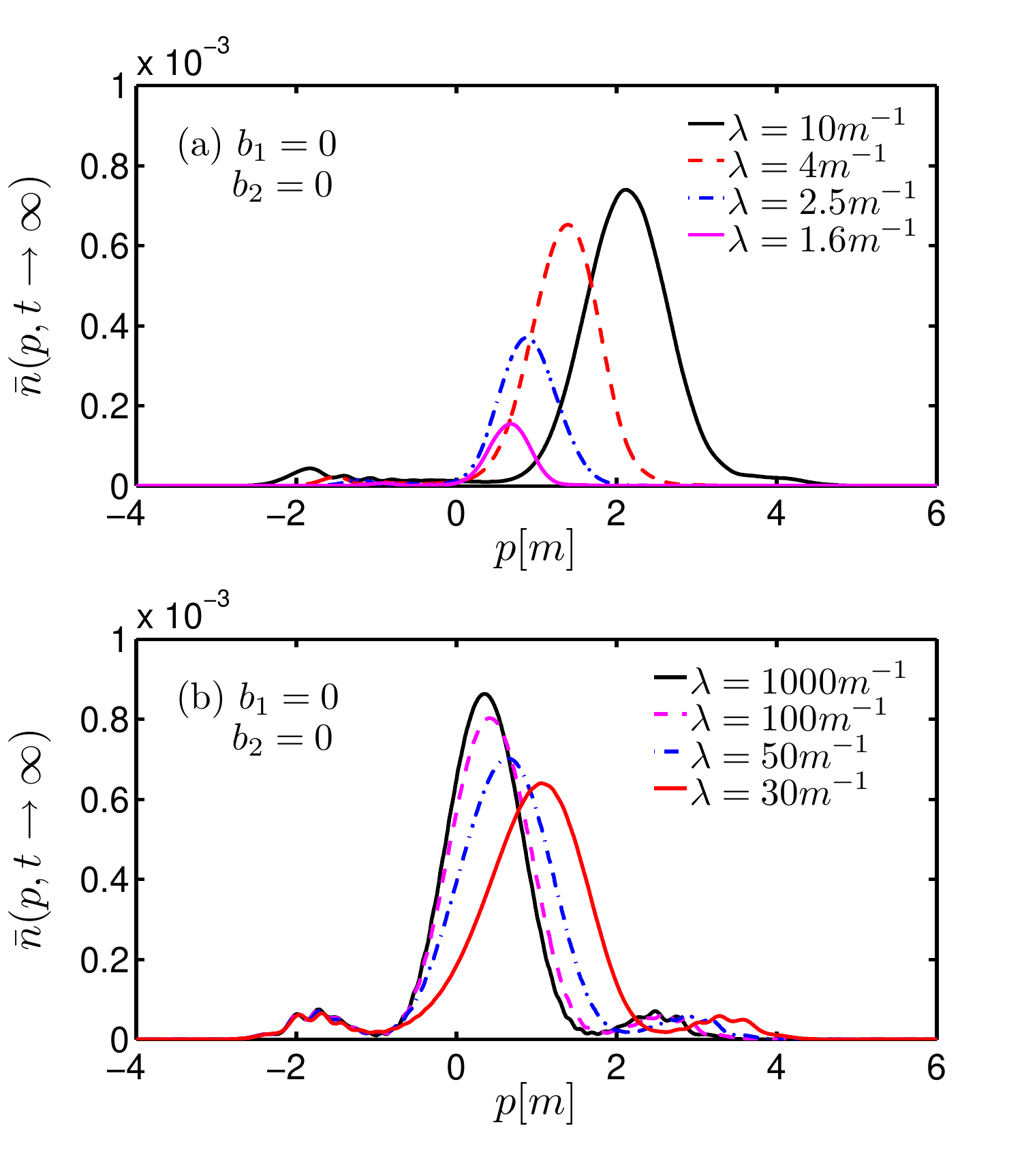}
\caption{(color online). Reduced momentum spectrum for different spatial scales for two-color field
$E\left(x,t\right)$ = $E_{1s}\left(x,t\right)$ +$E_{2w}\left(x,t\right)$
without chirp $b_{1}=b_{2}=0$. The field parameters are given in Eq. (\ref{Fixed parameters}).}
\label{fig:5}
\end{figure}

We will investigate the pair production for two-color field
$E\left(x,t\right)$ = $E_{1s}\left(x,t\right)$ +$E_{2w}\left(x,t\right)$ without chirp,
that is, the chirp parameters $b_{1}=b_{2}=0$.
The results for the momentum spectrum at small and
large spatial scales are presented in Fig. \ref{fig:5}(a) and (b), respectively.
Compared with the results of a single field as shown in Fig. \ref{fig:1} and Fig. \ref{fig:3},
the number density of created $e^{+}e^{-}$ pair increases significantly
because of the dynamically assisted Sauter-Schwinger mechanism \cite{Orthaber:2011cm, Ababekri:2019dkl}.
It is noted that the momentum spectrum peaks obtained by $E\left(x,t\right)$ without chirp are smaller
than that in the single field $E_{2w}\left(x,t\right)$ with the maximum chirp $b_{2}=0.9 \omega_{2}/\tau$
in the last section (Fig. \ref{fig:3}(d)) which indicates that the chirp parameter plays an important role in the momentum distribution.

It can also be seen from the Fig. \ref{fig:5}(a), the generation of $e^{+}e^{-}$ pair does not tend to terminate when the spatial
scales decreases further ($\lambda \leq 10 m^{-1}$), because the contribution from the high frequency field
does not decrease much by the finite spatial scale.

\subsection{Two-color field with chirping only for $E_{1s}\left(x,t\right)$}\label{LF}

\begin{figure}[htbp]\suppressfloats
\includegraphics[scale=0.55]{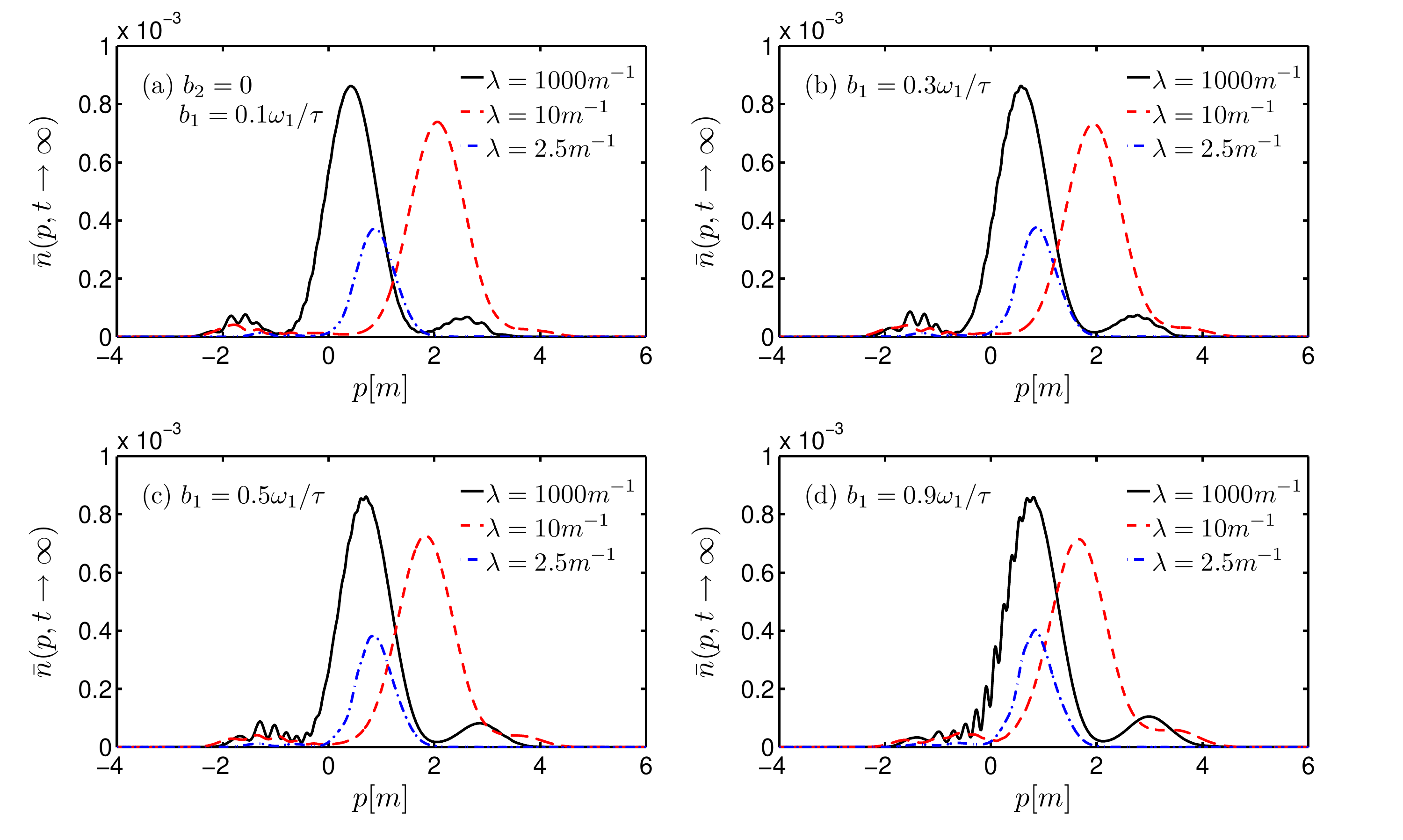}
\caption{(color online). Reduced momentum spectrum for different spatial scales for
two-color field
$E\left(x,t\right)$ = $E_{1s}\left(x,t\right)$ +$E_{2w}\left(x,t\right)$.
The chirping is only for $E_{1s}\left(x,t\right)$ with
$b_{1} [\omega_{1}/\tau]$ as $0.1$, $0.3$, $0.5$ and $0.9$, from (a) to (d) respectively.
The fields parameters are given in Eq. (\ref{Fixed parameters}).}
\label{fig:6}
\end{figure}

In the case of two-color field $E\left(x,t\right)$ =$E_{1s}\left(x,t\right)$ +$E_{2w}\left(x,t\right)$
with chirping only for $E_{1s}\left(x,t\right)$, i.e.,
$b_1=0.1 \omega_{1}/\tau, 0.3 \omega_{1}/\tau, 0.5 \omega_{1}/\tau, 0.9 \omega_{1}/\tau$ and $b_2=0$,
the reduced momentum spectrum is shown in Fig. \ref{fig:6}.
As can be seen in Fig. \ref{fig:6}(a) and (b), the small chirp has a little effect on the momentum spectrum as explained in Sec. \ref{result E1s}.
However, when the chirp parameter increases, such as $b_1=0.5 \omega_{1}/\tau$ and $b_1=0.9 \omega_{1}/\tau$,
the momentum spectrum signatures change significantly, especially in
the quasi-homogeneous ($\lambda=1000 m^{-1}$) case, which
can be seen from Fig. \ref{fig:6}(c) and Fig. \ref{fig:6}(d).
It can be seen that the momentum spectrum has been significantly changed,
even though the height is almost constant.
Moreover, the effects of chirp parameters can be understood in more detail
by obtaining the total particle yield and enhancement factor for various chirp values as a function of spatial scale $\lambda$.

\begin{figure}[htbp]\suppressfloats
\includegraphics[scale=0.42]{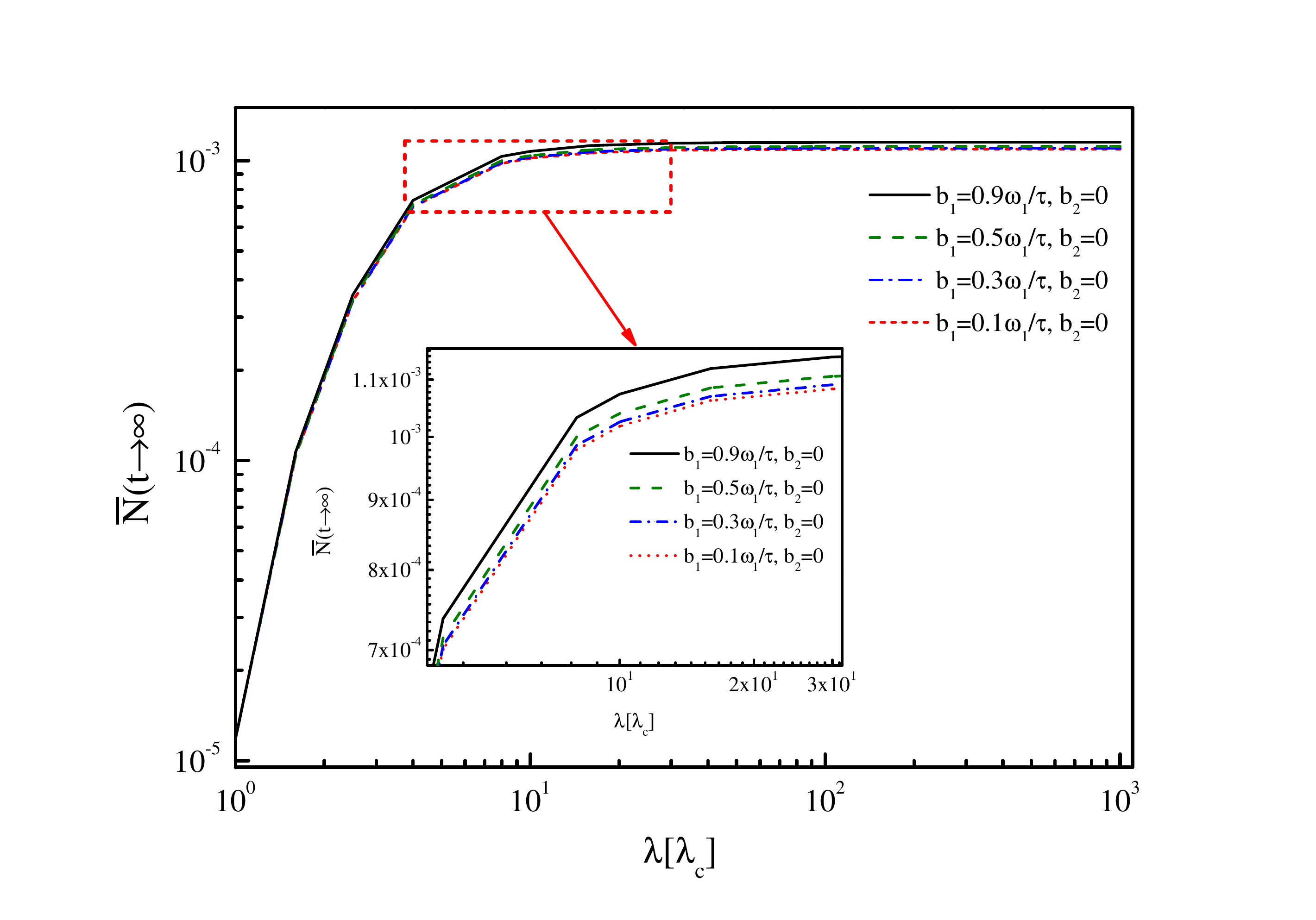}
\caption{(color online). Reduced total yield $\bar N (t \rightarrow \infty)$
for two-color field
$E\left(x,t\right)$ = $E_{1s}\left(x,t\right)$ +$E_{2w}\left(x,t\right)$ with
different chirp as a function of the spatial scales.
The chirping is only for $E_{1s}\left(x,t\right)$.
The parameters are the same as in Fig. \ref{fig:6} and the insert is the enlargement of finite spatial scales.}
\label{fig:7}
\end{figure}

Fig. \ref{fig:7} shows the total yield of created particles as a function of the spatial scales for
different chirp parameters.
It can be seen that for each chirp parameter, the total yield of created $e^{+}e^{-}$ pair in the
quasi-homogeneous region ($\lambda  \textgreater 100 m^{-1}$)
basically remains a constant reflecting the uniform approximation for the combined field.
However, for the small spatial scale, the total yield drops rapidly as the field energy for strong and fast pulse is also decreased.
Thus, the result is understandable and indicates that there is a slight improvement in the total yield
with the increase of chirp value. The same numerical result for finite spatial scale ($\lambda  \textless 4 m^{-1}$)
is given more clearly in the insert of Fig. \ref{fig:7}.

\begin{figure}[htbp]\suppressfloats
\includegraphics[scale=0.42]{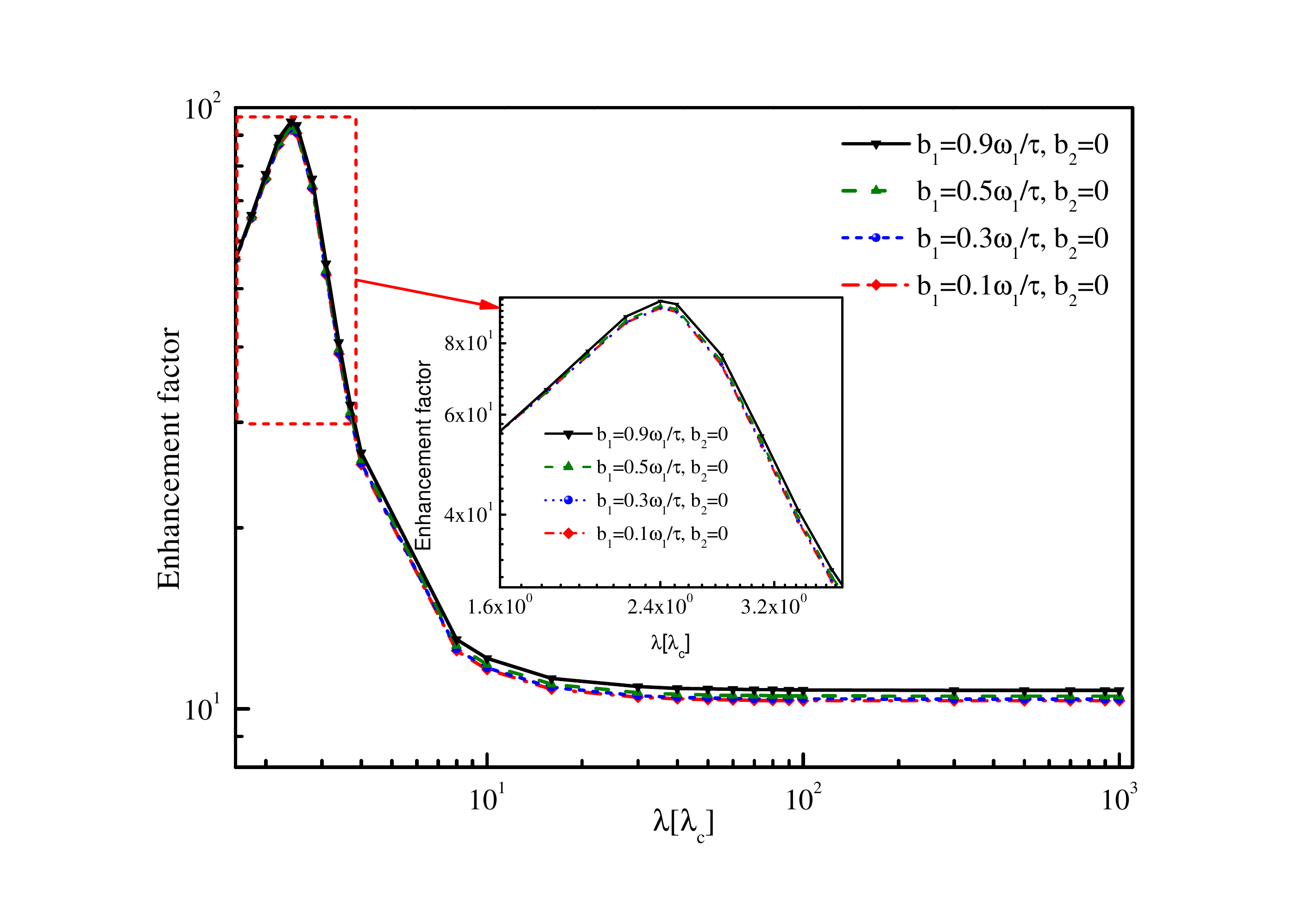}
\caption{(color online). Enhancement factor of the reduced total yield $\bar N_{1s+2w} / (\bar N_{1s}+\bar N_{2w})$
for different chirp as a function of the various spatial scales.
The parameters are the same as in Fig. \ref{fig:6}.}
\label{fig:8}
\end{figure}

In order to investigate whether there is an optimal chirp value or/and spatial scale for $e^{+}e^{-}$ pair
generation for two-color field with chirping only for $E_{1s}\left(x,t\right)$,
we further explore the enhancement factor of particle production for
different chirp values as a function of the spatial scale.
The enhancement factor is defined as the ratio $\bar N_{1s+2w} / (\bar N_{1s}+\bar N_{2w})$.
In this expression, $\bar N_{1s+2w}$ is the reduced total yield
of particles created for combined electric fields $E\left(x,t\right)$=$E_{1s}\left(x,t\right)$ +$E_{2w}\left(x,t\right)$
with chirping only for $E_{1s}\left(x,t\right)$ (chirp of $b_{1}$ [$\omega_{1}/\tau$] as $0.1$, $0.3$, $0.5$ and $0.9$),
and $\bar N_{1s}+\bar N_{2w}$ denote the direct summation of reduced total yield
for strong fields and weak fields with the certain chirp parameters.
We obtain that the enhancement factor of the total yield for different chirp values, as shown in Fig. \ref{fig:8}.

From Fig. \ref{fig:8}, we can see that when the chirp parameter is given, the enhancement factor decreases
with the decrease of the spatial scales in the extremely small spatial scale range $1.6 m^{-1} \textless \lambda \textless 2.4 m^{-1}$, and it increases with the decrease of the spatial scale
at the range of $ 2.4 m^{-1} \textless \lambda \textless 100 m^{-1}$.
For $\lambda  \textgreater 100 m^{-1}$, the enhancement factor is almost a constant for different chirp parameters,
which means that the effects of chirp parameter on the enhancement factor in quasi-homogeneous space are almost negligible.
However, we can get that the enhancement factor increases with
the increase of chirp values at a certain spatial scale.
It is noted that the chirp parameter has a little effect on the
total particle number of the various spatial scales and the momentum spectrum when $\lambda = 2.5 m^{-1}$ and $\lambda = 10 m^{-1}$,
but it has a visible effect on the momentum spectrum when the spatial width increases to $\lambda=1000 m^{-1}$, especially in the case of
the large chirp value $b_1=0.9 \omega_{1}/\tau$.

From the numerical results obtained above and what we have discussed, we can obtain that $\lambda = 2.4 m^{-1}$ and $b_{1}=0.9 \omega_{1}/\tau$
are the optimal values of the spatial scale and chirp to get the largest enhancement factor of created particles in this subsection.

\subsection{Two-color field with chirping only for $E_{2w}\left(x,t\right)$}\label{HF}

\begin{figure}[htbp]\suppressfloats
\includegraphics[scale=0.5]{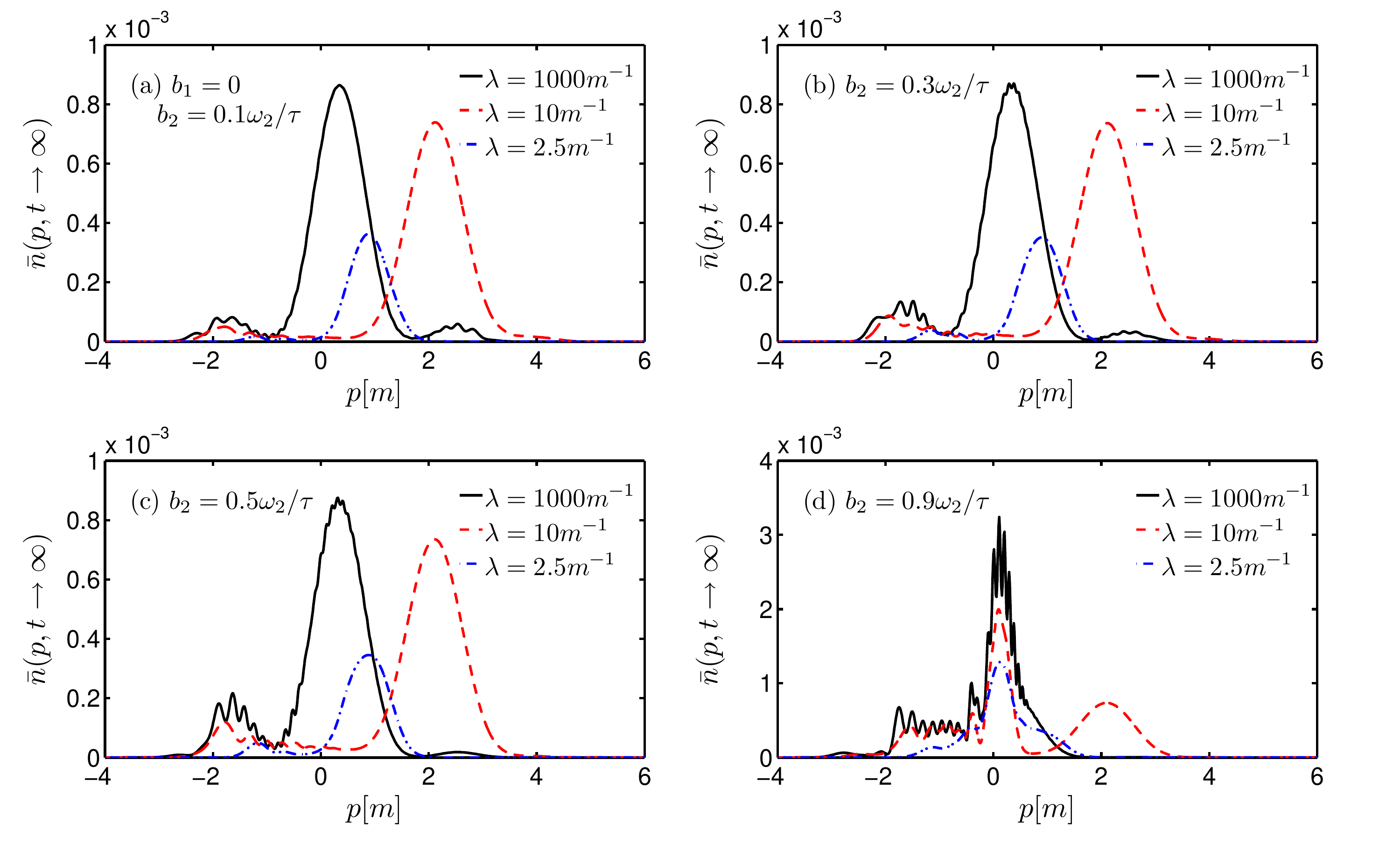}
\caption{(color online). Reduced momentum spectrum for different spatial scales for two-color field
$E\left(x,t\right)$ = $E_{1s}\left(x,t\right)$ +$E_{2w}\left(x,t\right)$.
The chirping is only for $E_{2w}\left(x,t\right)$ with
$b_{2}[\omega/\tau]$ as $0.1$, $0.3$,
$0.5$ and $0.9$, from (a) to (d) respectively.
The field parameters are given in Eq. (\ref{Fixed parameters}).}
\label{fig:9}
\end{figure}

For two-color field $E\left(x,t\right)$=$E_{1s}\left(x,t\right)$ +$E_{2w}\left(x,t\right)$
with chirping only for $E_{2w}\left(x,t\right)$, i.e.,
the chirp parameters $b_{2}=\alpha \omega_{2}/\tau$ with $\alpha=0.1$, $0.3$, $0.5$, $0.9$, the reduced momentum spectrum is exhibited in Fig. \ref{fig:9}.
The model of the field is also given as Eq. (\ref{FieldMode}), while the chirp parameters $b_{1}=0$.
Compared with the results of a single weak field $E_{2w}\left(x,t\right)$,
it can be seen that the momentum spectrum in the
present electric field model is more complex and the range of spectrum expands in momentum space.
The complicated reduced momentum spectrum for various spatial scales is due to the presence of
interference effect. Most importantly, the peak value of momentum spectrum increases with chirp,
which is also understood by the dynamically assisted Sauter-Schwinger mechanism.

It can be seen in Fig. \ref{fig:9}(a),
the oscillation distribution around the main peak of the momentum spectrum has a little difference
in the quasi-homogeneous case ($\lambda =1000 m^{-1}$),
when the chirp is small as $b_2 = 0.1 \omega_{2}/\tau=0.0014 m^{2}$.
For the intermediate chirp parameters $b_2 = 0.3 \omega_{2}/\tau=0.0042 m^{2}$ and $b_2 = 0.5 \omega_{2}/\tau=0.007 m^{2}$,
the oscillation of the momentum spectra become complex, see Fig. \ref{fig:9}(b) and (c).
When the chirp is large as $b_2 = 0.9 \omega_{2}/\tau$, the momentum spectrum has significant interference signatures, as shown in Fig. \ref{fig:9}(d).
For small space scales $\lambda =2.5 m^{-1}$ and $\lambda =10 m^{-1}$,
the effect of small chirp on the momentum spectrum is not obvious.
These responses of the momentum spectrum
to the chirp parameter means that the interference effect becomes stronger with the increase of chirp and
can also be understood  qualitatively with the interference
between the complex conjugate pairs of turning points, see Sec. \ref{SemiDis}.
Meanwhile, one sees that the height of peak is enhanced by at least four times
from $b_{2}=0.1 \omega_{2}/\tau$ to $b_{2}=0.9 \omega_{2}/\tau$.

\begin{figure}[htbp]\suppressfloats
\includegraphics[scale=0.42]{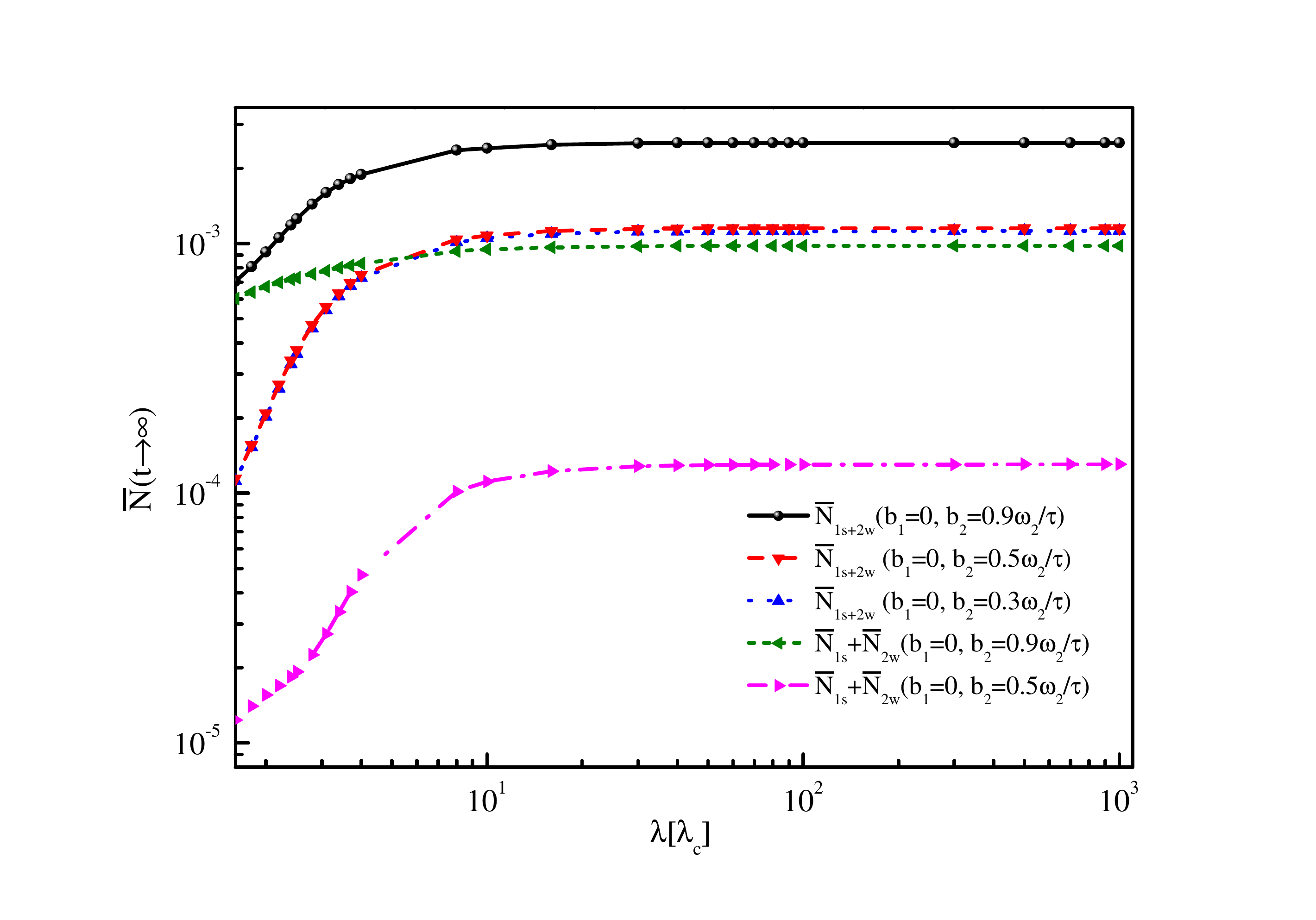}
\caption{(color online). Reduced total yield as a function of the spatial scales for different setups.
$\bar N_{1s+2w}$ is the total yield of two-color field
$E\left(x,t\right)$ = $E_{1s}\left(x,t\right)$ +$E_{2w}\left(x,t\right)$.
The chirping is only for
$E_{2w}\left(x,t\right)$ with $b_{2}=0.9 \omega_{2}/\tau$ (solid line), $0.5 \omega_{2}/\tau$ (dashed line)
and $0.3 \omega_{2}/\tau$ (dotted line).
$\bar N_{1s}+ \bar N_{2w}$
is simple direct summation of two individual field contribution for two kinds of $b_{2}$ [$\omega_{2}/\tau$]
as $0.9$ (dashed dotted line)
and $0.5$ (short dashed line).}
\label{fig:10}
\end{figure}

It is well known that the dynamically assisted electric field has a great influence on the generation of $ e^{+}e^{-} $ pair
including the momentum spectrum and the total particle number \cite{Orthaber:2011cm}.
Therefore, in Fig. \ref{fig:10} we consider that the reduced total yield of created particles as a function of the spatial scale
for different setups. Note that the total yield $\bar N_{1s+2w}$ is obtained by considering $E\left(x,t\right)$ = $E_{1s}\left(x,t\right)$ +$E_{2w}\left(x,t\right)$
and the chirping is only applied to $E_{2w}\left(x,t\right)$ with $b_{2}=0.9 \omega_{2}/\tau$, $0.5 \omega_{2}/\tau$ and $0.3 \omega_{2}/\tau$, respectively. As a comparison the yield $\bar N_{1s}+ \bar N_{2w}$ is obtained by a direct summation of individual yield of each field
$E_{1s}\left(x,t\right)$ with $b_{1}=0$ and $E_{2w}\left(x,t\right)$ with $b_{2}=0.9 \omega_{2}/\tau$ and $b_{2}=0.5\omega_{2}/\tau$, respectively.

From the dashed line and dotted line in the Fig. \ref{fig:10}, we can see the total yield increases slightly when
the chirp parameter varying from  $b_{2}=0.3 \omega_{2}/\tau$ to $b_{2}=0.5 \omega_{2}/\tau$.
However compared with the chosen maximum chirp at $b_{2}=0.9 \omega_{2}/\tau$
as shown by the solid line, the total yield in the latter case increases by at least $2$ times.
We think that the role of effective frequency increasing greatly when the chirp is large would not only provide more energy to created particles but also create more particles. In particular, it is observed that
$\bar N_{1s}+ \bar N_{2w}$ ($b_{2}=0.9\omega_{2}/\tau$) \textgreater $\bar N_{1s+2w}$ ($b_{2} \le 0.5 \omega_{2}/\tau$)
at the finite spatial scale $\lambda \textless 6 m^{-1}$. This result indicates that the high frequency field is extremely sensitive to the chirp, meanwhile,
the choice of the appropriate chirp is also important.

\begin{figure}[htbp]\suppressfloats
\includegraphics[scale=0.42]{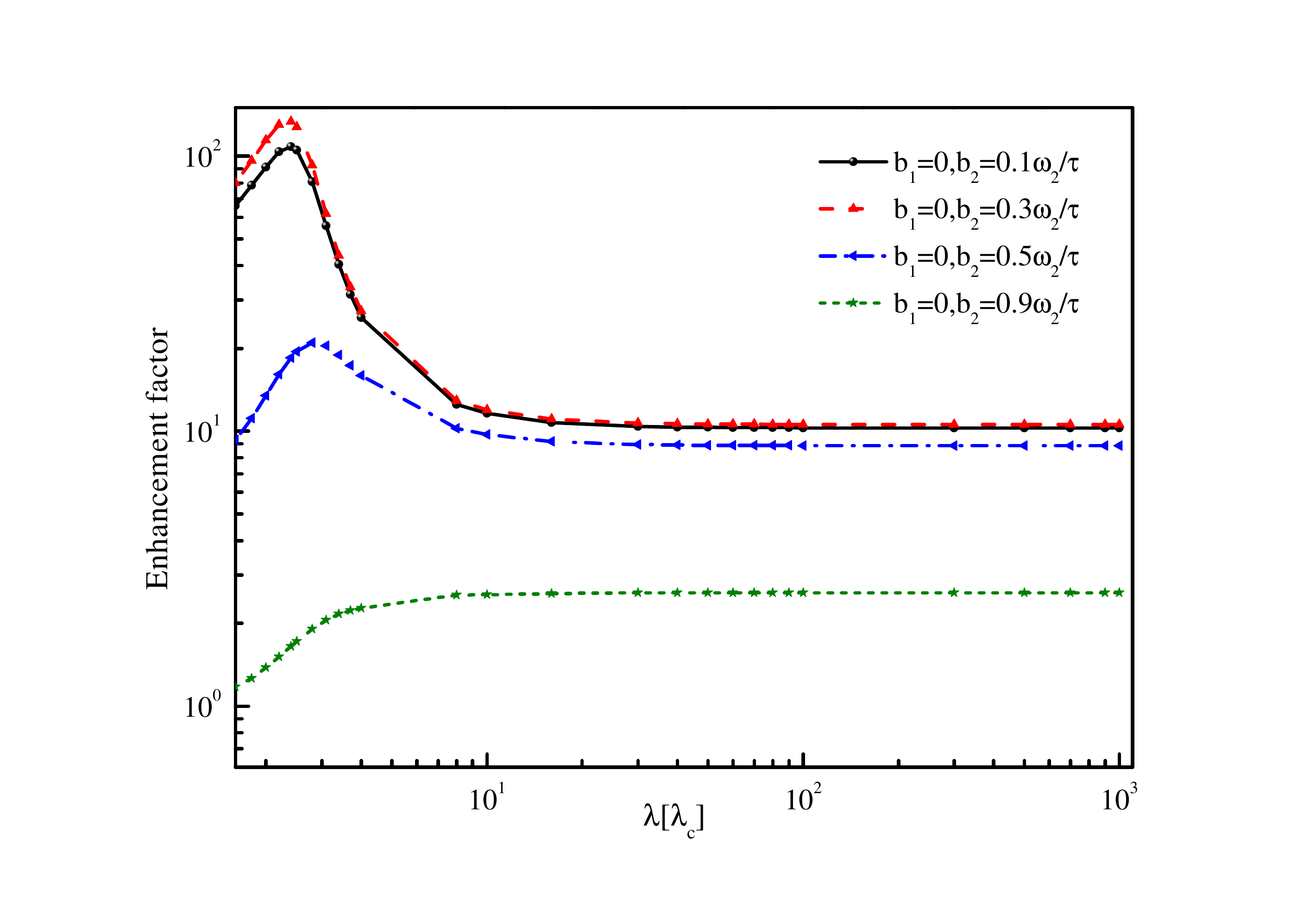}
\caption{(color online). Enhancement factor of the total number $\bar N_{1s+2w} / (\bar N_{1s}+\bar N_{2w})$ for different chirp values as a function of the various spatial scales.
The parameters are the same as in Fig. \ref{fig:9}.}
\label{fig:11}
\end{figure}

We also investigate the enhancement factor of total yield
for different chirp values as a function of the various spatial scales
and the result is shown in Fig. \ref{fig:11}. It is obvious that the enhancement factor is
almost a constant for each chirp in the quasi-homogeneous region $\lambda \geq 1000 m^{-1}$.
Moreover, the variation of enhancement factor is complex in the finite spatial scale.
For relative small chirp values $b_{2}\leq 0.3 \omega_{2}/\tau$, the enhancement factor increases
at the spatial scale $2.4 m^{-1} \textless \lambda \textless 100 m^{-1}$ and
decreases at the spatial scale $1.6 m^{-1} \textless \lambda \textless 2.4 m^{-1}$ with the shrinking of the spatial scale.
Thus, we obtain that the enhancement factor reaches maximum at the extremely spatial scale $\lambda = 2.4 m^{-1}$.
As can be seen from the solid line ($b_{2}=0.1\omega_{2}/\tau$)
and dashed line ($b_{2}=0.3\omega_{2}/\tau$) in Fig. \ref{fig:11},
the larger the chirp parameter is, the larger the enhancement factor.
For the larger chirp value $b_{2}=0.5\omega_{2}/\tau$  and $b_{2}=0.9\omega_{2}/\tau$,
enhancement factor is smaller than that with the small chirp values.
These results can also be qualitatively obtained in Fig. \ref{fig:10}.

\subsection{Two-color field with chirping for both $E_{1s}\left(x,t\right)$ and $E_{2w}\left(x,t\right)$}\label{AF}

\begin{figure}[htbp]\suppressfloats
\includegraphics[scale=0.5]{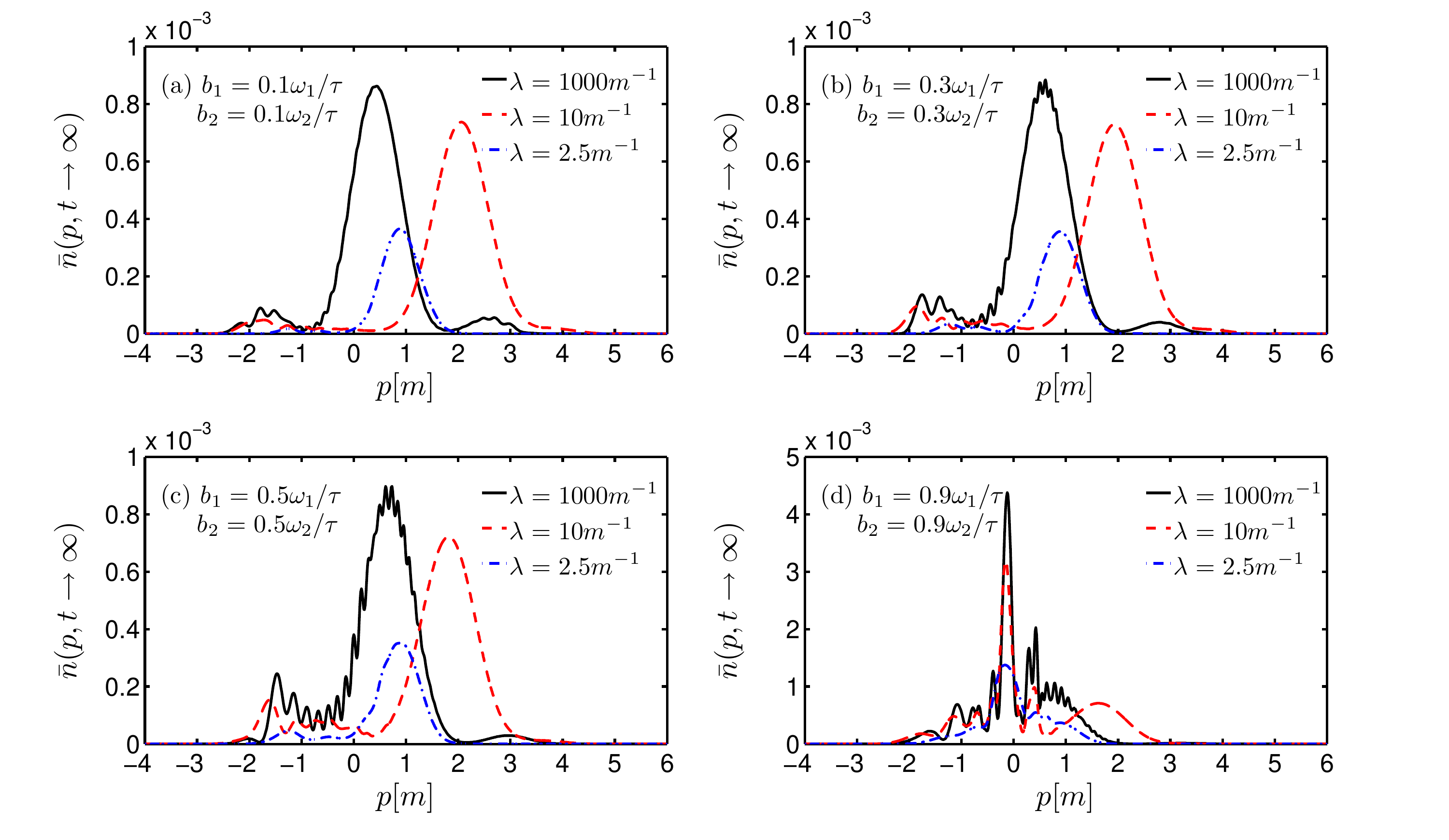}
\caption{(color online). Reduced momentum spectrum for
two-color field $E\left(x,t\right)$ = $E_{1s}\left(x,t\right)$ +$E_{2w}\left(x,t\right)$
with chirping for both fields.
The chirps are $b_{1} [\omega_{1}/\tau]$ as $0.1$, $0.3$, $0.5$, $0.9$
and $b_{2} [\omega_{2}/\tau]$ as $0.1$, $0.3$, $0.5$, $0.9$.
The fields parameters are given in Eq. (\ref{Fixed parameters}).}
\label{fig:12}
\end{figure}

In the case of two-color field $E\left(x,t\right)$=$E_{1s}\left(x,t\right)$ +$E_{2w}\left(x,t\right)$
with chirping for both $E_{1s}\left(x,t\right)$ and $E_{2w}\left(x,t\right)$,
the model of the external field is given in Eq. (\ref{FieldMode}) with $b_1 \neq 0 $ and $b_{2} \neq 0$.
The reduced momentum spectrum for various spatial scales is shown in Fig. \ref{fig:12}.
From Fig. \ref{fig:12}(a) and (b) one can see that the weak
interference pattern appears in the momentum spectrum.
For $\lambda=1000m^{-1}$, the large chirp causes a complex oscillation which can be understood
as enhancing of the interference effect of created particles shown in Fig. \ref{fig:12}(c).
In case of $b_1 = 0.9 \omega_{1}/\tau$ and $b_2 = 0.9 \omega_{2}/\tau$ in Fig. \ref{fig:12}(c),
we can see that some additional peaks appear beside the main peak in the momentum spectrum.
More importantly, the peak of the momentum spectrum in this case is higher compared
to the results of the other cases.

\begin{figure}[htbp]\suppressfloats
\begin{center}
\includegraphics[scale=0.42]{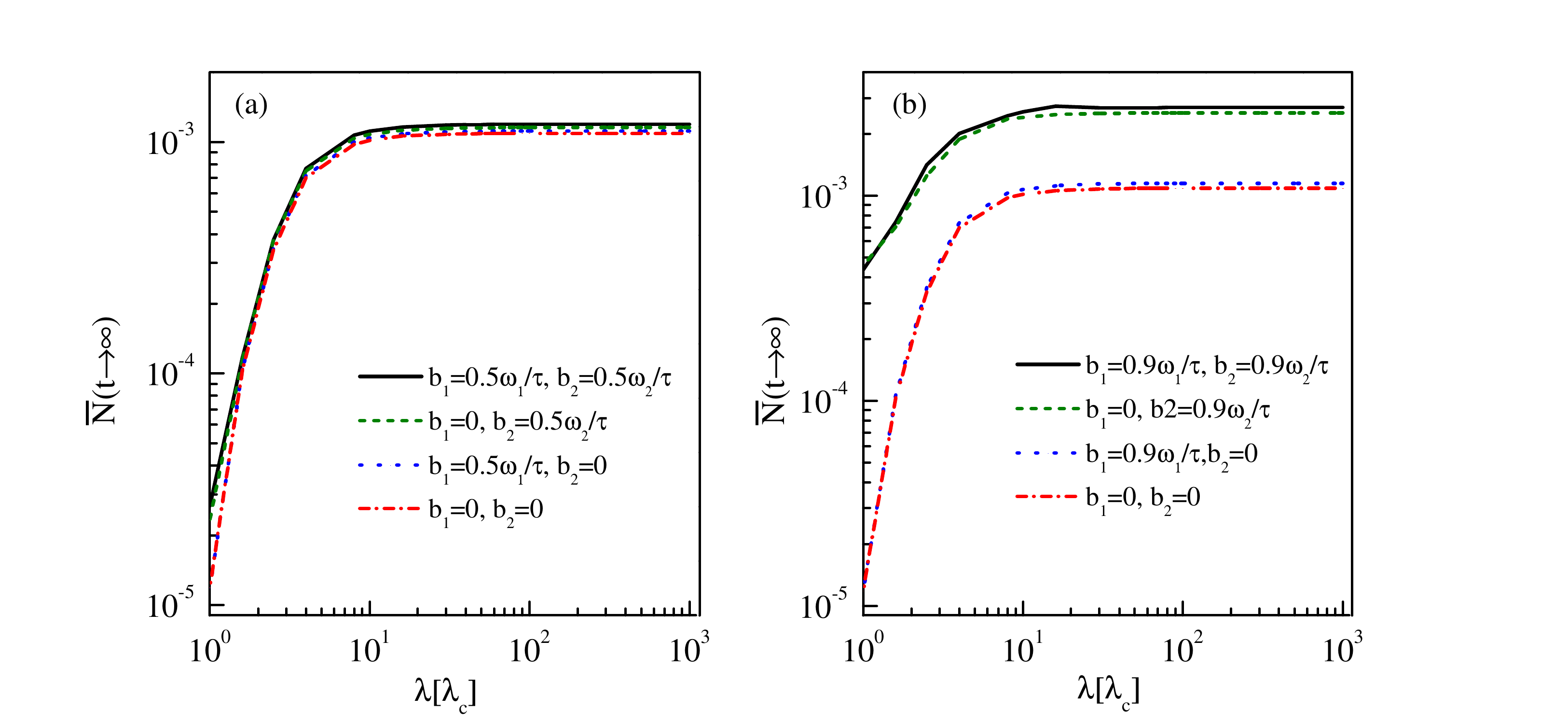}
\end{center}
\caption{(color online). Reduced total yield as a function of the various spatial scales
for two-color field $E\left(x,t\right)$ = $E_{1s}\left(x,t\right)$ +$E_{2w}\left(x,t\right)$
with four different chirping cases, $b_{1}=0, b_{2}=0$ (red dashed dotted line),
$b_{1}\neq 0, b_{2}=0$ (blue dotted line),
$b_{1}=0, b_{2}\neq0$ (green dashed line),
and $b_{1}\neq0, b_{2}\neq0$ (black solid line).
The nonzero chirp parameters $b_{i}$ [$\omega_{i}/\tau$], ($i=1,2$), are chosen as $0.5$
in ($a$) and $0.9$ in ($b$).
The other parameters are given in the text.}
\label{fig:13}
\end{figure}

Fig. \ref{fig:13} shows the reduced total yield of created particles
as a function of the spatial scale for
two-color field $E\left(x,t\right)$ = $E_{1s}\left(x,t\right)$ +$E_{2w}\left(x,t\right)$
with four different chirping cases,
$b_{1}\neq0, b_{2}\neq0$ (black solid line),
$b_{1}=0, b_{2}\neq0$ (green dashed line),
$b_{1}\neq 0, b_{2}=0$ (blue dotted line),
and $b_{1}=0, b_{2}=0$ (red dashed dotted line).
In Fig. \ref{fig:13}(a), the nonzero chirp parameters are $b_{i}=0.5\omega_{i}/\tau$, where $i=1,2$, however, in
Fig. \ref{fig:13}(b) they are chosen as $0.9$.

From Fig. \ref{fig:13}(a), it can be seen that for relatively small nonzero chirp parameters, the total yield
has small differences at a certain spatial scale in the presented four cases.
From Fig. \ref{fig:13}(b), we can see that the reduced total yield increases very much
when the chirp value is chosen large as $b_{1}=0.9 \omega_{1}/ \tau$ and $b_{2}=0.9 \omega_{2}/ \tau$.
When $\lambda \geq 40 m^{-1}$, the total yield tends to an almost flat line.
In the field $E\left(x,t\right)$ = $E_{1s}\left(x,t\right)$ +$E_{2w}\left(x,t\right)$,
the reduced total yield does not increase too much when the chirp acts only on $E_{1s}\left(x,t\right)$,
compared with the case without chirp.
However, the reduced total yield increases greatly when the chirp acts on both of the fields
or only on $E_{2w}\left(x,t\right)$, compared with the case that the chirp acts only on $E_{1s}\left(x,t\right)$.
Specifically, the reduced total yield
increases more than one order of magnitude in extremely small spatial scales $\lambda \approx 1.6 m^{-1}$
and increases by $2.38$ and $2.2$ times in quasi-homogeneous space region $\lambda \geq 100 m^{-1}$.

\begin{figure}[htbp]\suppressfloats
\begin{center}
\includegraphics[scale=0.42]{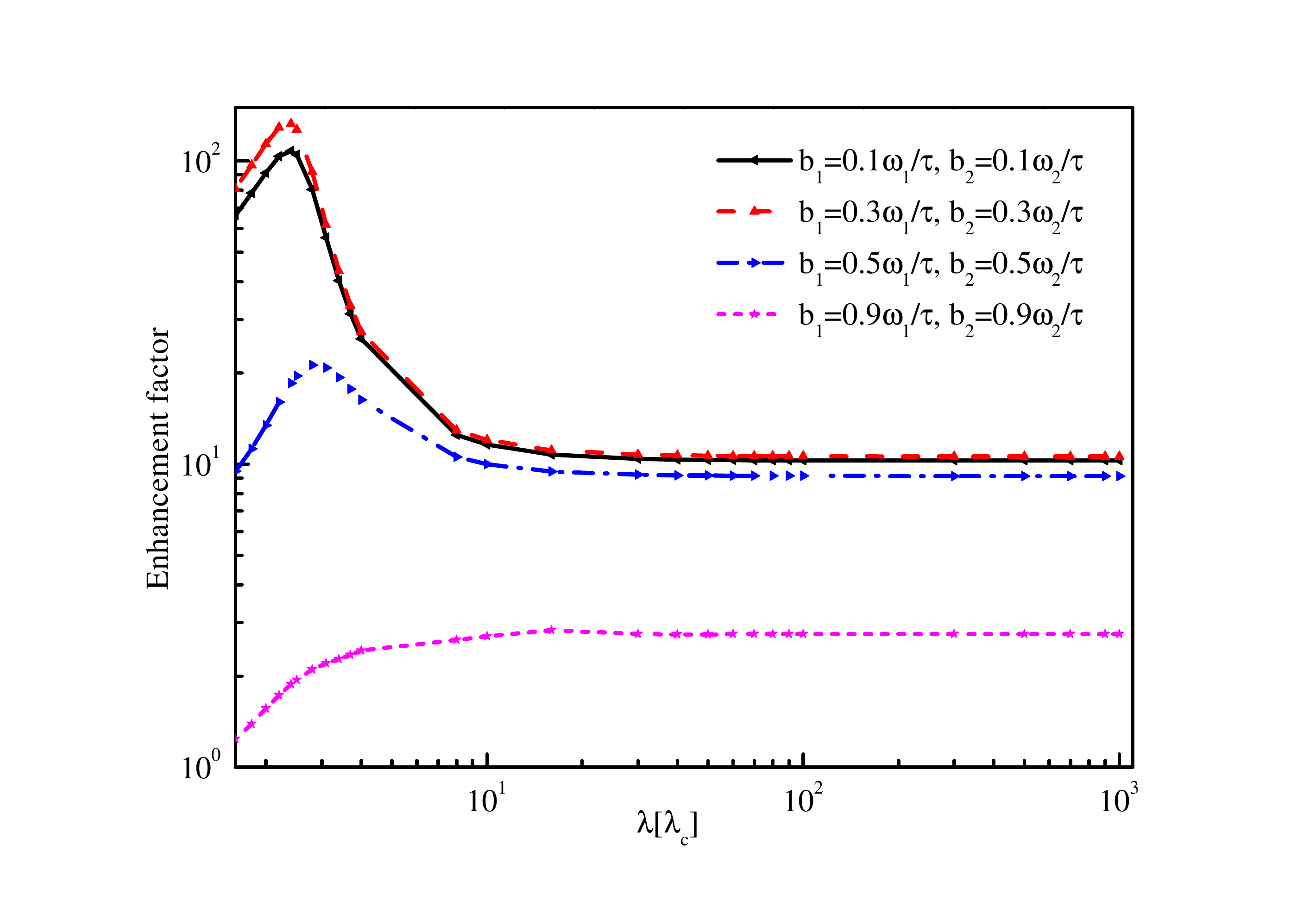}
\end{center}
\caption{(color online). Enhancement factor of the total number $\bar N_{1s+2w} / (\bar N_{1s}+\bar N_{2w})$
for different chirp values as a function of the various spatial scales.
The parameters are the same as in Fig. \ref{fig:12}.}
\label{fig:14}
\end{figure}

Fig. \ref{fig:14} shows the enhancement factor of the total yield
for different chirp values as a function of the various spatial scales.
It is found that the enhancement factor is
almost a constant for each chirp in the quasi-homogeneous region.
However, the variation of enhancement factor is complex in the finite spatial scale.
We can also obtain that the enhancement factor is the largest at $\lambda = 2.4 m^ {-1}$ for the small
chirp values $b_{i}[\omega_{i}/\tau]\leq 0.3,(i=1,2)$ and at $\lambda = 16 m^ {-1}$
for the largest chirp value.

\begin{table}[htbp]
\caption{Optimal spatial scales related to the optimal enhancement factor
at the chosen optimal chirp parameters.}
\centering
\begin{ruledtabular}
\begin{tabular}{lccc}
Different chirping & Chirp parameter ($\omega_{i}/\tau$), ($i=1,2$) & Spatial scale ($m^ {-1}$)& Enhancement factor\\
\hline
$b_{1} \neq 0$, $b_{2}=0$ &$b_{1}=0.9 $ & $\lambda = 2.4 $& $94.756$ \\
$b_{1} = 0$, $b_{2}\neq0$ &$b_{2}=0.3$  &$\lambda = 2.4 $ & $133.584$ \\
$b_{1}\neq0$, $b_{2}\neq0$ &$b_{1}=b_{2}=0.3$ & $\lambda = 2.4$ &$132.517$\\
\end{tabular}
\end{ruledtabular}
\vskip12pt
\label{table:1}
\end{table}

We collect the optimal chirp parameters and spatial scales related to the optimal enhancement factors from
Fig. \ref{fig:8}, Fig. \ref{fig:11} and Fig. \ref{fig:14}. The corresponding chirping scheme and optimal values are given in Table \ref{table:1}.
It is found that the enhancement factor is always the optimal value at the extremely small spatial scale $\lambda = 2.4 m^ {-1}$
for different optimal chirp parameters in the study regime.
We also find that the enhancement factor is the largest
when the chirp acts only on high frequency field $E_{2w}\left(x,t\right)$ (second row in the table).
Compare the result with the first row (a large chirp parameter acts on the low frequency field in the combinational
field), we find that the effect of large chirp parameter in
a low frequency field is not significant. Interestingly,
in third row (a small chirp parameter acts on both of the fields in the combinational
field), the enhancement factor is smaller than the one in the second row. The physical mechanism behind the phenomenon
needs to study further.

\subsection{Semiclassical analysis and discussion}\label{SemiDis}

In order to see why the interference effect of the momentum spectrum we obtained and why the pairs yield is increased by the chirping in some cases, we employ the semiclassical WKB approach to make some discussion on obtained numerical results in this subsection. In fact the physical picture is related to the quantum tunneling problem in an effective scattering potential of applied background fields, which is reflected in the typical structure of the turning points.

We know that the WKB approach is a quasiclassical one to study the potential scattering problem so that it seems to be more effective in the Schwinger tunneling regime $\gamma \ll 1 $, however, it has been reasonably extended to perturbative multiphoton regime in
many works \cite{Popov1968QuasiclassicalAF,Brezin:1970xf,2012PhDT........77D,Gong:2019sbw,Ababekri:2019qiw}. In fact as early as in around of 1970's the WKB associated to the turning points has been worked for the study of atomic ionization \cite{Popov1968QuasiclassicalAF} and also that of pair production \cite{Brezin:1970xf} for both of $\gamma \ll 1$ and $\gamma \gg 1$.
Moreover, recently Dumlu has made a series of study for the fields in the multiphoton regime \cite{2012PhDT........77D}. For example, he studied the field with $E_{0}=0.1 m^{2}$ and $\omega=0.5m$ and discussed the corresponding turning points distribution in detail. In fact the applicability of WKB and the corresponding turning points is seen by that
the pair creation rate can be expressed either as $N\approx \exp(-2K)$ \cite{Dumlu:2011rr} or as $N\sim \exp[-(\pi E_{cr}/2eE) g(\gamma)]$ , see Eq.(48) in Ref.\cite{Brezin:1970xf}, therefore, $K$ is associated to $K=-(\pi E_{cr}/2eE) g(\gamma)$. The difference for $\gamma \ll 1$ and $\gamma \gg 1$ is that
$g(\gamma) \simeq1$ when $\gamma \ll 1$ and $g(\gamma)=(4/\pi \gamma)\ln(2\gamma)+ o(1/\gamma)$ when $\gamma\gg 1$, just as the Eq. (49) of Ref.\cite{Brezin:1970xf}. In fact the method of stationary phase to get the pair production rate when $\gamma \gg 1$ in Ref.\cite{Brezin:1970xf} is equivalent to that by the WKB approach with complex turning points structure.

The approximate expression of the particle creation rate for multiple turning points is (refer to Ref.\cite{Dumlu:2011rr})
\begin{equation}\label{AE}
\begin{aligned}
N\approx \sum\limits_{t_{i}}e^{-2K_{i}}- \sum\limits_{t_{i}\neq t_{j}} 2\cos(2\theta_{(i,j)})
e^{-K_{i}-K_{j}},
\end{aligned}
\end{equation}
with
$$ K_{i}=\left\vert \int_{t_{i}^{*}}^{t_{i}}{w_{\bm p}(t)dt} \right\vert, $$
and
$$\theta_{(i,j)}=\left\vert\int_{Re (t_{i})}^{Re(t_{j})}{w_{\bm p}(t)dt} \right\vert,$$
where $t_{i}$ and $t_{j}$ are the solutions of equation $w_{\bf p}(t)= \sqrt{m^{2}+p_{\perp}^{2}+(p_{x}-eA(t))^{2}}=0$, i.e., the turning points and $\rm{Re}(t_i)$ is the real part of point. Note that the turning points closest to the $t$ real axis tend to dominate the pair production arising from the first term in Eq.(\ref{AE}), and the oscillatory behavior of the momentum spectrum, i.e., interference effect is due to the interference terms $\theta_{(i,j)}$ in Eq. (\ref{AE}).

\begin{figure}[htbp]
  \centering
  \subfigure{\includegraphics[width=4cm]{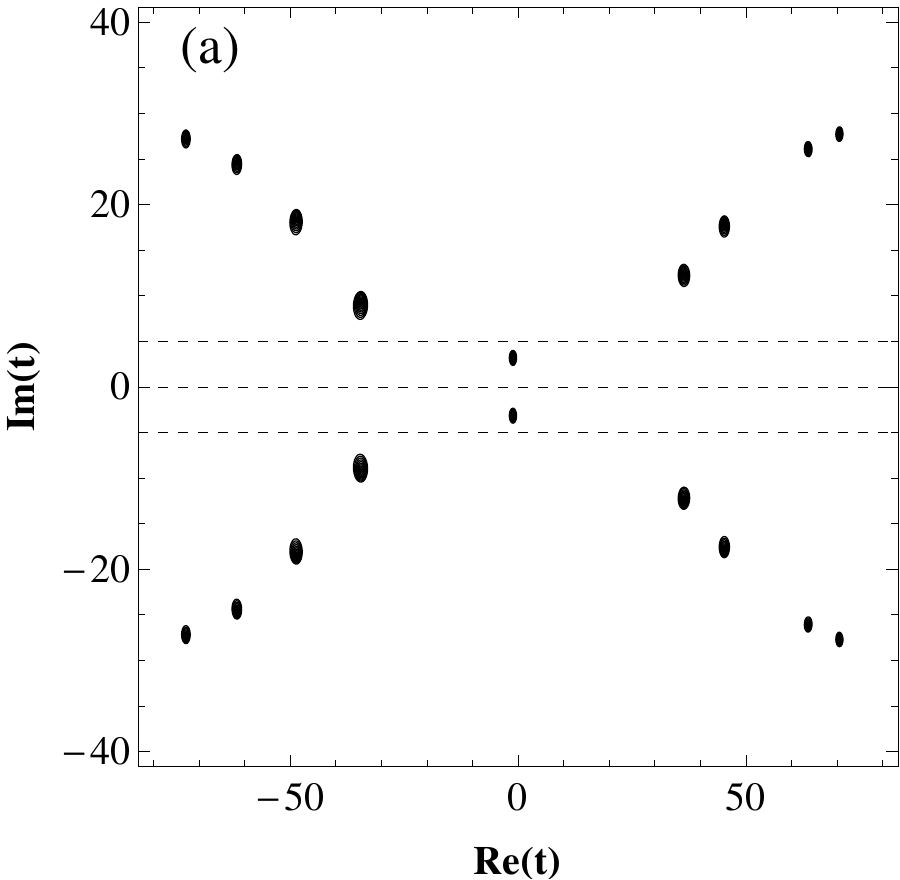}}
  \hspace{6.5mm}
  \subfigure{\includegraphics[width=4cm]{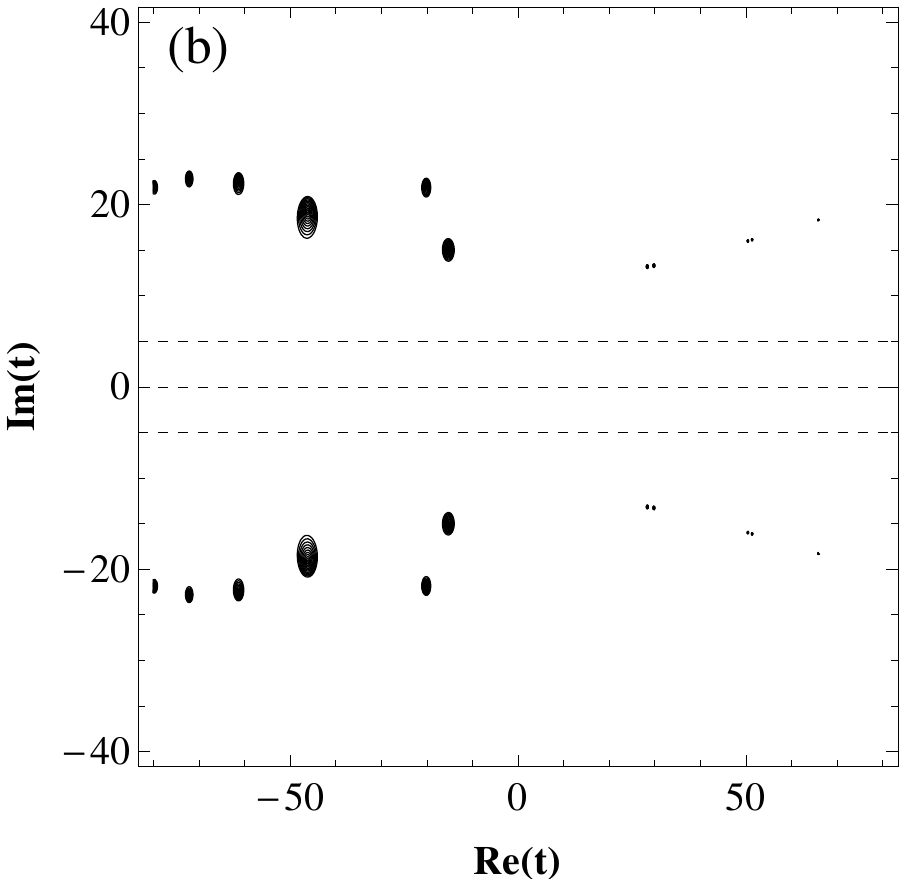}}
  \hspace{6.5mm}
  \subfigure{\includegraphics[width=4cm]{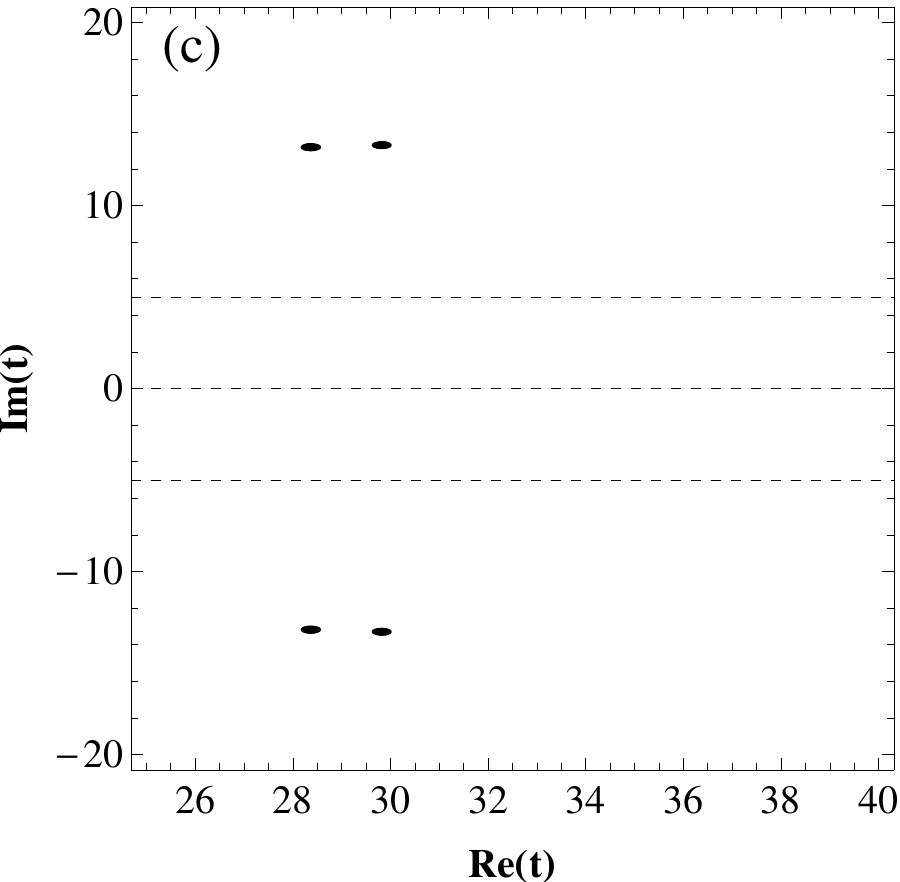}}
  \caption{ Contour plots of $|w_{\bm p}(t)|^{2}$ in the complex $t$ plane,
  showing the location of turning points for $E_{1s}$ where $w_{\bm p}(t)=0$.
  These plots are for the vanishing momentum $p_{x}=0$.
  Form (a) to (b) the chirp values are $b_{1}=0$ and $b_{1}=0.9\omega_{1}/\tau$.
  Panel (c) is the amplified plot of the two set of pairs of turning points at around $\rm{Re}(t)\sim30$ in panel (b).
  The other field parameters are the same as in Fig. \ref{fig:1}.
  The three dashed lines at $\rm{Re}(t)=0$ and adjacent the real $t$ axis are for eyes guidance.}
  \label{fig:15}
\end{figure}

\begin{figure}[htbp]
  \centering
  \subfigure{\includegraphics[width=6cm]{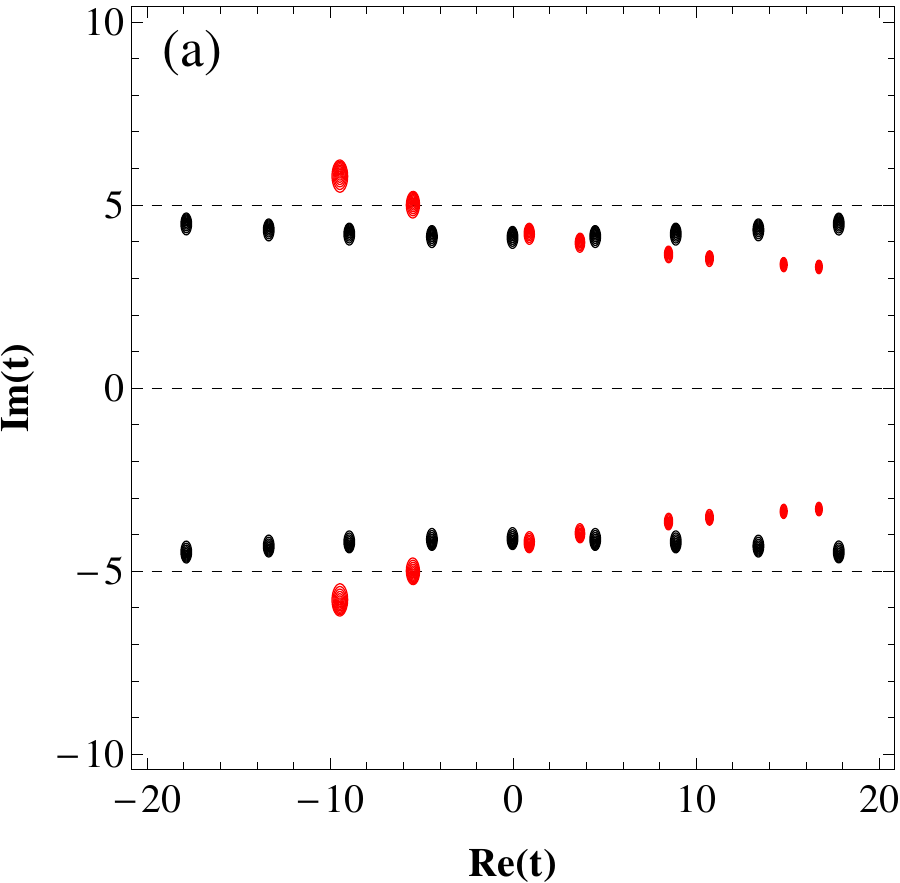}}
  \hspace{7mm}
  \subfigure{\includegraphics[width=6cm]{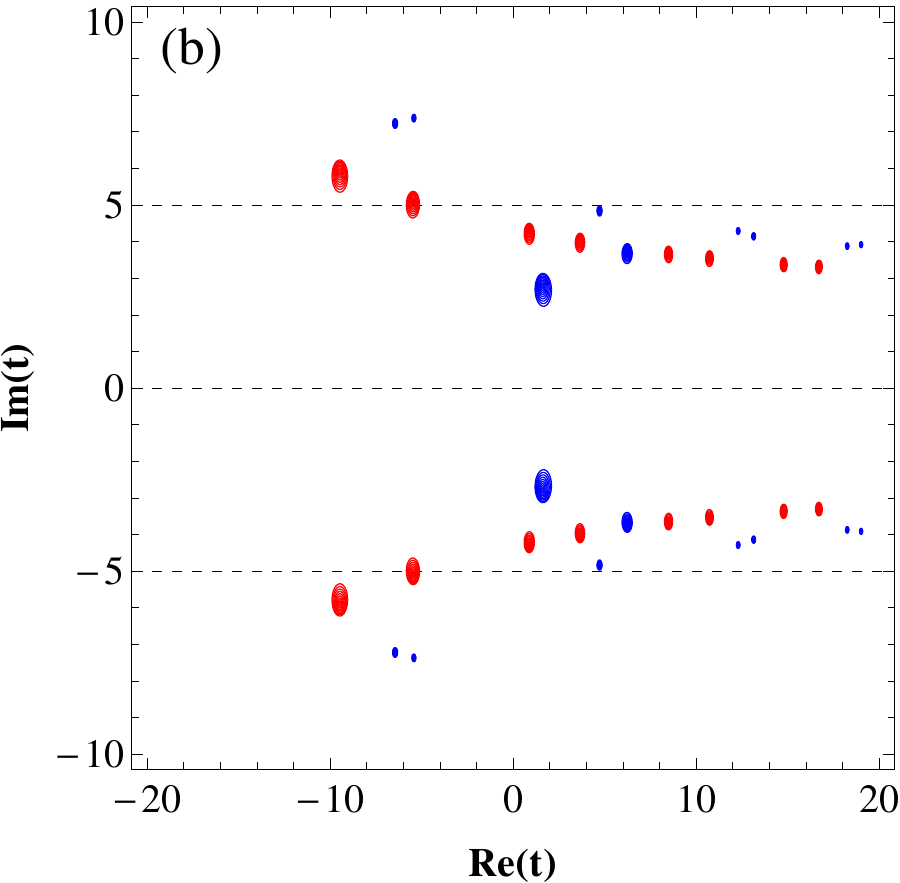}}
  \caption{(color online). Contour plots of $|w_{\bm p}(t)|^{2}$ in the complex $t$ plane,
  showing the location of turning points for different models of external field where $w_{\bm p}(t)=0$.
  These plots are for the vanishing momentum $p_{x}=0$.
  Panel (a): The plot is for $E_{2w}$, with $b_{2}=0$ (black) and
  $b_{2}=0.9\omega_{2}/\tau$ (red).
  Panel (b): The red dots are for $E_{2w}$ (same as that in panel (a)),
  and the blue dots
  for $E(x,t)=E_{1s}+E_{2w}$ with chirping only for $E_{2w}$, i.e., $b_{1}=0$, $b_2=0.9\omega_{2}/\tau$. The other field parameters are given in Eq. (\ref{Fixed parameters}). The three dashed lines at $\rm{Re}(t)=0$ and adjacent the real $t$ axis are for eyes guidance.}
  \label{fig:16}
\end{figure}

Since the physical reasons are similar we just choose three representative situations, i.e., (i) strong field $E_{1s}$ without and with chirping; (ii) weak filed $E_{2w}$ without and with chirping; and (iii) combined fields $E(x,t)=E_{1s}+E_{2w}$ with chirping only for $E_{2w}$, to discuss and understand the numerical results we have obtained above.
The corresponding typical turning point structures are illustrated in Fig. \ref{fig:15} in case of (i) and Fig. \ref{fig:16} in cases of (ii) and (iii), respectively. To discuss turning points structure clearly, we depict them at the vanishing momentum $p_{x}=0$.

For the case of $E_{1s}$ without chirp $b_{1}=0$, in the Fig. \ref{fig:15}(a),
one can see that there is only a single dominant pair of turning point located the center of figure,
thus there is no interference. This result can be used to qualitatively explain the corresponding
momentum spectrum, as shown in Fig. \ref{fig:1}(a).
When chirp value $b_{1}=0.9\omega_{1}/\tau$, the distribution of turning points
seems to be more complicated as shown in Fig. \ref{fig:15}(b),
we find that there are two dominant complex conjugate pairs of turning points
located in the right real $t$ regime, which can be clearly seen in Fig. \ref{fig:15}(c). And they have almost equidistant from the real axis. It is well known that the interference effect is strongest when the pairs of different turning points are at a comparable distance from the real axis, therefore, the obvious interference patterns appear in the momentum distribution shown in Fig. \ref{fig:1}(d). Moreover, in this chirping case, the two set of pairs of turning points at around $\rm{Re}(t)\sim30$ have almost the same distances from real axis at $|\rm{Im}(t)|=|\rm{Im}(t^{*})|\approx15$.
For the single pair of turning point without chirping at around $\rm{Re}(t)\sim 0$, the
distances from real axis at $|\rm{Im}(t)|=|\rm{Im}(t^{*})|\approx5$. These turning points characteristics indicates two facts that are qualitatively consistent with the numerical results of Fig. \ref{fig:1}(a) and (d) for $b_1=0$ and $b_{1}=0.9\omega_{1}/\tau$ under an approximate homogeneous case of $\lambda=1000m^{-1}$. On one hand, the closer distance to real axis means that the particles creation rate in the no-chirping case is large comparable to the chirping one. Indeed $n\left(p_{x}=0, t\rightarrow \infty\right)=4.67\times 10^{-5}m$ when $b_{1}=0$ is a little larger than $n\left(p_{x}=0, t\rightarrow \infty\right)=3.43\times 10^{-5}m$ when $b_{1}=0.9\omega_{1}/\tau$. On the other hand, because the interference effect is associated to the distances of turning points along the real-axis direction, so the neighbouring two set of pairs in Fig. \ref{fig:15}(c) implies an obvious interference effect which is revealed in the oscillation behaviour at around of $p=0$ in momentum spectrum, see Fig. \ref{fig:1}(d).

In Fig. \ref{fig:16}(a), the black dots are turning points of $E_{2w}$ without chirping,
and the red dots are turning points of $E_{2w}$ with large chirp $b_{2}=0.9\omega_{2}/\tau$. Obviously the turning points located in the positive real $t$ regime is much more close to real axis as well as more number in the case of chirping (red) compared to the case of no-chirping (black). This means that the particle number density is increased due to the closer of turning points to real $t$ axis and the stronger interference due to the appearance of more turning points. It is why  the stronger interference effects can be observed on the momentum spectrum as shown in Fig. \ref{fig:3} (d).

As a further comparison, we add $E_{1s}$ to $E_{2w}$ so that the model is a two-color field , the turning points structure is plotted in Fig. \ref{fig:16}(b) with $b_1=0$ and still $b_2=0.9\omega_{2}/\tau$, see the blue dots. From Fig. \ref{fig:16}(b) one can see that, compared with the one-color chirping field $E_{2w}$ (red dots), for the two-color field, even if the $E_{1s}$ has no chirping but the $E_{2w}$ chirping is kept, the turning points structure is changed to be that the dominant one-pair turning point is closer to $t$ real axis very much and also more points appeared in the right. Therefore, the appearance of more obvious interference effects in Fig. \ref{fig:9}(d) is understandable compared with that in Fig. \ref{fig:3} (d).

Certainly the other momentum spectra can also be understood by the similar discussions through the corresponding turning points structure. Since the physical picture as the semiclassical WKB approximation is simple and valid, it is not necessary to give a repeat presentation for more situations. What should be pointed out here is that the delicate understanding of the momentum signature and pairs yield beyond the WKB picture may be possible and it is worthy to be studied furthermore in future \cite{Kaminski:2018ywj}.

\section{Conclusion and outlook}\label{conclusion}

Within the DHW formalism, we investigated the reduced momentum spectrum, the
reduced total yield of the created pairs for either low or high frequency one-color field with different chirp values.
Meanwhile, we studied chirped dynamically assisted pair production for two-color
combinational fields, and the enhancement factor is obtained in the assisted case.
Moreover, we employ the semiclassical WKB approach and the corresponding turning points structure
to make some qualitative discussion.

For the strong low frequency field,
the momentum spectrum has almost the same peak value and oscillates significantly when the
chirp value is the largest, especially in the quasi-homogeneous region.
In the case of the high frequency weak field, when the chirp increases, the oscillatory pattern of the
spectrum becomes more complicated as the presence of interference effect. While an incomplete interference is seen in the
spectrum when some appropriate chirp is used.
More importantly, the peak value of reduced momentum spectrum and the total yield
increase strongly with the increasing of chirp parameters.
We have examined Ref. \cite{Orthaber:2011cm,Olugh:2019nej} and meanwhile expanded the dynamically assisted field results in the sense of having chirping. We obtained that the total yield of produced pair is indeed enhanced significantly in some cases by appropriate chirp parameters. However, our studies here exhibit very delicate and subtle effects about the chirping and the field forms on the pair production problem.

For two-color field with chirping only for $E_{1s}\left(x,t\right)$,
we found that the momentum spectrum depends on the chirp parameters
to some extent, which is more obvious in the quasi-homogeneous region. In addition,
the reduced total yield of created particles and the enhancement factor of total particle number
do not increase greatly with the increase of chirp parameter, which indicates that
the chirp parameter has a little effect on the pair production
in this case.
However, in the case of two-color field with chirping only for $E_{2w}\left(x,t\right)$,
the particle number density is sensitive to the chirp
and the reduced momentum spectrum shows a strong nonlinear feature.
The reduced total yield of particles in this electric field increases significantly,
which increases by at least $2$ times for the large chirp value and by one order of magnitude for the small chirp value,
compared with that in the case of a simple direct summation of two individual field.
It is also noted that, for $E\left(x,t\right)$ with chirping for both $E_{1s}\left(x,t\right)$ and $E_{2w}\left(x,t\right)$, strong interference effects are observed in the momentum distribution, so that the high total yield is achieved.
Specifically, in the chirped dynamically assisted scenario, when the chirps are acting on the two fields the reduced total yield is enhanced
by more than one order of magnitude in the field with a relative small spatial scale, while it is enhanced at least two times
in other case of field with larger spatial scales or even in the quasi-homogeneous region,
compared with that the chirp is acting only for the low frequency strong field.
Most importantly, we also obtain some optimal chirp parameters and spatial scales for enhancement factor of the total yield in different scenarios of the studied combined field.

We believe that these results presented in this paper are useful for having a better understanding about the
effect of the chirp parameter and provide further insight into the relation between
the structure of external field and pair creation.
Our study provides also the theoretical basis of some broader parameters range for possible experiments in the future by the modeling of
combinational fields that enhance the dynamically assisted pair production.

\begin{acknowledgments}
\noindent
We would like to thank Dr. M. Ababekri and N. Abdukerim for critical reading of the manuscript.
We are also grateful to Dr. Li Wang for useful discussion and valuable comments.
And we are thankful to the anonymous referee for helpful suggestions to improve the manuscript.
This work was supported by the National Natural Science Foundation of China (NSFC) under
Grant No. 11875007 and No. 11935008. The computation was carried out at the HSCC of the Beijing Normal University.

\end{acknowledgments}

\end{document}